\documentclass[twocolumn,aps,prc,superscriptaddress,showpacs,floatfix,longbibliography,nofootinbib]{revtex4-1}
\usepackage{url}
\usepackage{cancel}
\usepackage[colorlinks,linkcolor=blue,citecolor=blue,filecolor=black,urlcolor=blue]{hyperref}
\usepackage{epsfig,graphics}
\usepackage{graphicx}
\usepackage{dcolumn}
\usepackage{bm}
\usepackage[usenames]{color}
\usepackage{amssymb}
\usepackage{amsmath}
\usepackage{multirow}
\usepackage{float}
\usepackage{harpoon}
\usepackage{MnSymbol}
\usepackage{appendix}
\usepackage{color}
\usepackage{hyperref}
\usepackage{cleveref}

\begin{document}

\title{Deformation probes for light nuclei in their collisions at relativistic energies}
\author{Hai-Cheng Wang}
\affiliation{School of Physics Science and Engineering, Tongji University, Shanghai 200092, China}
\author{Song-Jie Li}
\affiliation{School of Physics Science and Engineering, Tongji University, Shanghai 200092, China}
\author{Lu-Meng Liu}
\affiliation{Physics Department and Center for Particle Physics and Field Theory, Fudan University, Shanghai 200438, China}
\author{Jun Xu}\email[Correspond to\ ]{junxu@tongji.edu.cn}
\affiliation{School of Physics Science and Engineering, Tongji University, Shanghai 200092, China}
\author{Zhong-Zhou Ren}\email[Correspond to\ ]{zren@tongji.edu.cn}
\affiliation{School of Physics Science and Engineering, Tongji University, Shanghai 200092, China}
\date{\today}
\begin{abstract}
We have investigated the performance of anisotropic flows $\langle v_n^2 \rangle$, transverse momentum fluctuations $\langle \delta p_T^2 \rangle $, and their correlations $\langle v_n^2 \delta p_T \rangle$ in central collisions at relativistic energies as probes of deformation parameters $\beta_n$ of colliding nuclei, if these nuclei are light nuclei with large $\beta_n$ and different configurations of $\alpha$ clusters. The effects from higher-order $\beta_n$ terms are illustrated by derived relations based on the overlap of two nuclei with uniform density distributions and by dynamic simulations of collisions of heavy nuclei whose density distributions are of a deformed Woods-Saxon (WS) form. While the linear relations between $\beta^2_n$, $\langle v_n^2 \rangle$, and $\langle \delta p_T^2 \rangle$ and that between $\beta^3_n$ and $\langle v_n^2 \delta p_T \rangle$ can be violated for extremely large $\beta_{n}$, they are mostly valid for realistic values of $\beta_n$, as long as the density distribution of colliding nuclei can be described by a deformed WS form. However, these linear relations are generally not valid with more realistic density distributions of light nuclei with $\alpha$ clusters, and the amount of deviation depends on the detailed $\alpha$-cluster configurations. Care must be taken when one tries to extract the deformation of light nuclei, and specific probes for $\alpha$-cluster structures in these nuclei are very much needed.
\end{abstract}
\maketitle

\section{introduction}

Understanding the density distribution in finite nuclei, especially their shapes, is a fundamental goal of nuclear physics. Comparing with traditional low-energy experiments, it has been realized recently that relativistic heavy-ion collisions can serve as an alternative way of achieving this goal (see, e.g., Ref.~\cite{Bally:2022vgo}). Compared to collisions of spherical nuclei, collisions of deformed nuclei provide more possible geometries and configurations, which may lead to different areas and shapes of the overlap region, and thus enhance the anisotropic collective flow and affect the transverse momentum spectrum~\cite{Filip:2009zz,Shou:2014eya,Giacalone:2019pca,Zhang:2021kxj,Jia:2021tzt,Jia:2021wbq,Jia:2021qyu,Jia:2021oyt,Giacalone:2021udy}. The situation becomes more clear in central collisions, where various probes have been proposed to extract the deformation of colliding nuclei, e.g., a linear correlation is observed between $\langle \epsilon_n^2 \rangle$ and $\beta_n^2$~\cite{Jia:2021tzt}, where $\epsilon_n$ represents the $n$th-order anisotropy coefficient of the overlap region and $\beta_n$ is the $n$th-order deformation parameter of the colliding nuclei. Since the initial anisotropy in coordinate space is eventually transformed into the final anisotropy in momentum space, the square of the final $n$th-order anisotropic flow $\langle v_n^2 \rangle$ is expected to be linearly correlated with $\beta_n^2$ as well. There are also other probes such as the transverse momentum fluctuation $\langle \delta p_T^2 \rangle$ originated from the fluctuation in the initial overlap area characterized by $\langle \delta d_\perp^2 \rangle$~\cite{Giacalone:2019pca,PhysRevLett.130.212302}, as well as the correlation between the anisotropic flow and the transverse momentum fluctuation~\cite{Jia:2021qyu}. These probes have been used to successfully extract the deformation parameters of $^{96}$Ru~\cite{Zhang:2021kxj}, $^{96}$Zr~\cite{Zhang:2021kxj}, $^{197}$Au~\cite{Giacalone:2021udy}, and $^{238}$U~\cite{STAR2024}. It was proposed that light nuclei such as $^{20}$Ne could have a huge deformation which can be measured through their collisions~\cite{Bally:2022vgo,Giacalone:2024luz}. The recent $^{16}$O+$^{16}$O collisions under analysis by the STAR Collaboration also provide a good opportunity to investigate the structure of light nuclei.

So far the linear relations between $\beta_n^2$ and $\langle \epsilon_n^2 \rangle$ ($\langle v_n^2 \rangle$) or $\langle \delta d_\perp^2 \rangle$ ($\langle \delta p_T^2 \rangle$)~\cite{Jia:2021qyu,Jia:2021tzt} work well generally based on studies by assuming that the density distribution of colliding nuclei is of a deformed WS form with small $\beta_n$. On one hand, whether the linear relation is valid for large values of $\beta_n^2$ such as those for light nuclei needs further investigations. On the other hand, it has been shown that light nuclei are generally formed by different configurations of $\alpha$ clusters~\cite{Brink:1970ufk,Bauhoff:1984zza,Freer:2017gip,Tohsaki:2017hen,BIJKER2020103735,Zhou:2019hor,Li:2020exz}, and their density distribution can no longer be described by a deformed Woods-Saxon form. The typical $3-\alpha$ cluster and $4-\alpha$ cluster structure in $^{12}$C and $^{16}$O, respectively, have been under hot investigation for a long time~\cite{Freer:2017gip,BIJKER2020103735,Tohsaki:2017hen}, and $\alpha$ clusters may have linear-chain and triangle configurations in $^{12}$C, and linear-chain, tetrahedron, square, and Y-shape configurations in $^{16}$O~\cite{Liu:2023gun}. $^{20}$Ne is largely deformed as proposed in Ref.~\cite{Bally:2022vgo}, and it also has a special internal structure with five $\alpha$ clusters~\cite{Zhou:2023vgv,Ropke:2024rlr}. Another candidate of light nuclei could be $^{10}$Be, which has a life time long enough to be used for heavy-ion experiments, and it could be composed of two $\alpha$ clusters and two valence neutrons~\cite{Myo:2023alz,Li:2023msp}. These different configurations of light nuclei may affect anisotropic flows in their collisions~\cite{PhysRevC.95.064904,PhysRevC.97.034912,Behera:2023nwj,PhysRevC.104.L041901,Broniowski:2013dia}. It is of interest to extract the deformation or the $\alpha$-cluster structure of these nuclei with proper probes in their collisions at relativistic energies.

The purpose of the present paper is to investigate whether the probes used to extract the deformation of colliding nuclei work for largely deformed nuclei and for light nuclei with different $\alpha$-cluster configurations. If not, how large is the error in extracting $\beta_n$ from these probes for nuclei whose density distribution can not be described by a deformed WS form. Section~\ref{theory} presents how the density distributions of $^{12}$C, $^{16}$O, $^{20}$Ne, and $^{10}$Be are obtained from the microscopic cluster model with Brink wave function~\cite{Brink:1970ufk}, and reviews briefly the structure of a multiphase transport (AMPT) model~\cite{Lin:2004en} used to simulate the collision dynamics of these nuclei at relativistic energies. Section~\ref{results} first illustrates the performance of the probes for the deformation of colliding nuclei by using $^{96}$Zr+$^{96}$Zr collisions, with the density distribution of $^{96}$Zr parameterized by a deformed WS form, and then investigate the difference in the probes using density distributions of a deformed WS form and more realistic $\alpha$-cluster configurations for light nuclei. We conclude and outlook in Sec.~\ref{summary}.

\section{theoretical framework}
\label{theory}

In this section, we will first present the framework of calculating the density distributions of $^{12}$C, $^{16}$O, $^{20}$Ne, and $^{10}$Be with $\alpha$-cluster structures, and then briefly review the AMPT model used for simulating the collisions of these light nuclei at relativistic energies. The deformation probes for colliding nuclei in their central collisions at relativistic energies will be discussed, and numerical relations related to these probes with simple assumptions will be derived.

\subsection{A microscopic cluster model}

In order to obtain the density distribution of light nuclei, we adopt the following Hamiltonian
\begin{eqnarray}
  \hat{H} &=&
  \sum^A_{i=1} \ E_i^{\textup{cm}}
  + \sum_{i<j} \ V^{NN}({r}_{ij}) \notag\\
  &+& \sum_{i<j} \ V^{Cou}({r}_{ij}) + \sum_{i<j} \ V^{ls}({r}_{ij}).
  \label{eq:Hamiltonian}
\end{eqnarray}
The summation in the above Hamiltonian is over the total nucleon number $A$. The first term represents the kinetic energy in the center-of-mass (c.m.) frame, the second term is Volkov No.2 force~\cite{Volkov_1965_NP} representing the effective nucleon-nucleon interaction, the third term is the Coulomb interaction, and the fourth term is the spin-orbit interaction, with $\vec{r}_{ij}$ being the relative coordinates between nucleon $i$ and nucleon $j$. The form of the effective nucleon-nucleon interaction can be expressed as
\begin{equation}
  V^{NN}({r}_{ij})
  = ( V_1 e^{-\alpha_1 {r}^2_{ij}} - V_2 e^{-\alpha_2 {r}^2_{ij}} )
  ( W - M \hat{P}_{\sigma} \hat{P}_{\tau} + B \hat{P}_{\sigma} - H \hat{P}_{\tau} ),
  \label{eq:Volkov}
\end{equation}
where $\hat{P}_{\sigma}$ and $\hat{P}_{\tau}$ are the spin and isospin exchange operator, respectively, and $V_1 = -60.650 \ \textup{MeV}$, $V_2 = 61.140 \ \textup{MeV}$, $\alpha_1 = 0.980 \ \textup{fm}^{-2}$, and $\alpha_2 = 0.309 \ \textup{fm}^{-2}$ are used in all calculations. For the calculation of $^{12}$C, $^{16}$O, and $^{10}$Be, we use $W = 0.4$, $M = 0.6$, and $B = H = 0.125$, determined from the phase shift data of $\alpha$-nucleon and $\alpha-\alpha$ scatterings as well as the binding energy of deuteron~\cite{Volkov_1965_NP, Itagaki_2000_PRC}, and they are also used in Ref.~\cite{Lyu:2015ika}. For the calculation of $^{20}$Ne, we use $W = 0.38$, $M = 0.62$, and $B = H = 0$, as used in Ref.~\cite{Itagaki:2010ha}. The G3RS force is
used for the spin-orbit interaction with the form
\begin{equation}
V^{ls}({r}_{ij}) = u_{ls}(e^{-\alpha_3 {r}^2_{ij}} - e^{-\alpha_4 {r}^2_{ij}}) \vec{L} \cdot \vec{S} \hat{P}_{31},
\end{equation}
with $u_{ls}=1600$ MeV, $\alpha_3=5$ fm$^{-2}$, and $\alpha_4=2.778$ fm$^{-2}$ as in Ref.~\cite{Lyu:2015ika}, and $\hat{P}_{31}$ being the operator which projects the two-nucleon system to the $(S=1, T=1)$ state.

For $^{12}$C, $^{16}$O, and $^{20}$Ne, their ground states can be described with the Bloch-Brink wave function, namely, the creation operator $C^{\dagger}_{\alpha}$ of $\alpha$ clusters acting on the vacuum state $| \mathrm{vac} \rangle$, i.e.,
\begin{equation}
  \begin{aligned}
      | \Phi^{\mathrm{Brink}} \rangle = (C^{\dagger}_{\alpha})^n
      | \mathrm{vac} \rangle,
  \end{aligned}
  \label{eq:12C}
\end{equation}
with $n=A/4$ being the number of $\alpha$ clusters, and
\begin{equation}
  \begin{aligned}
      C^{\dagger}_{\alpha}
      & = \int d^3 {r}_1 \cdots d^3 {r}_4 \\
      & \times
      \phi(\vec{r}_1 - \vec{R}) a^{\dagger}_{\sigma_1, \tau_1}(\vec{r}_1)
      \cdots \phi(\vec{r}_4 - \vec{R}) a^{\dagger}_{\sigma_4, \tau_4}(\vec{r}_4).
  \end{aligned}
  \label{eq:Brink-cluster_creation}
\end{equation}
In the above, the spatial part of a single particle has a Gaussian form
\begin{equation}
      \phi(\vec{r} - \vec{R}) \propto  \exp\left[- \frac{(\vec{r} - \vec{R})^2}{2 b^2}\right],
  \label{eq:single_particle}
\end{equation}
where the Gaussian width $b$ is fixed to be 1.46 fm for all nucleons~\cite{Itagaki_1995_PTP}, and $a^{\dagger}_{\sigma, \tau}$ is the creation operator for nucleon with spin $\sigma$ and isospin $\tau$.

For $^{10}$Be, we follow the framework in Ref.~\cite{Lyu:2015ika}, where the ground state is described by the Tohsaki-Horiuchi-Schuck-R\"opke (THSR) wave function, i.e.,
\begin{equation}
  \begin{aligned}
      | \Phi^{\mathrm{THSR}}_{^{10}\mathrm{Be}} \rangle
      = (C^{\dagger}_{\alpha})^2 (c^{\dagger}_{n})^2
      | \mathrm{vac} \rangle
      ,
  \end{aligned}
  \label{eq:10Be}
\end{equation}
where $C^{\dagger}_{\alpha}$ and $c^{\dagger}_{n}$ are the creation operators of $\alpha$ clusters and valence neutrons, respectively. The creation operator $C^{\dagger}_{\alpha}$ of $\alpha$ clusters in the THSR framework is reformulated as
\begin{equation}
    \begin{aligned}
        C^{\dagger}_{\alpha}
        & = \int \ d \vec{R}\ \mathcal{G}_{\alpha}(\vec{R})
        \int d^3 {r}_1 \cdots d^3 {r}_4 \\
        & \times
        \phi(\vec{r}_1 - \vec{R}) a^{\dagger}_{\sigma_1, \tau_1}(\vec{r}_1)
        \cdots \phi(\vec{r}_4 - \vec{R}) a^{\dagger}_{\sigma_4, \tau_4}(\vec{r}_4),
    \end{aligned}
    \label{eq:cluster_creation_operator}
\end{equation}
according to the $\alpha$-condensate picture~\cite{Zhou:2013ala}. Here a Gaussian container $\mathcal{G}_{\alpha}(\vec{R})$ is used to confine the motion of $\alpha$ clusters, i.e.,
\begin{equation}
    \begin{aligned}
        \mathcal{G}_{\alpha}(\vec{R})
        & = \exp\left(
            -\frac{{R}_x^2 + {R}_y^2}{\beta_{\alpha,xy}^2}
        -\frac{{R}_z^2}{\beta_{\alpha,z}^2}\right),
    \end{aligned}
    \label{eq:Gaussian_cluster}
\end{equation}
where $\beta_{\alpha,xy}$ and $\beta_{\alpha,z}$ are parameters to be optimized by the variational principle in the calculation.
The creation operator $c^{\dagger}_{n}$ of valence neutrons is formulated as
\begin{equation}
    \begin{aligned}
        c^{\dagger}_{n} =
        \int \ d \vec{R}\ \mathcal{G}_n(\vec{R})
        \int d^3 {r} \phi(\vec{r} - \vec{R}) a^{\dagger}_{\sigma, \tau}(\vec{r}).
    \end{aligned}
    \label{eq:valence_neutron_creation_operator}
\end{equation}
In the above, $\mathcal{G}_n(\vec{R})$ is the Gaussian container for valence neutrons, which is modulated by a phase factor~\cite{Lyu:2015ika} to reproduce the negative parity of the $\pi$-orbit, i.e.,
\begin{equation}
    \begin{aligned}
        \mathcal{G}_n(\vec{R})
        & = \exp\left(
        -\frac{{R}_x^2 + {R}_y^2}{\beta_{n,xy}^2}
        -\frac{{R}_z^2}{\beta_{n,z}^2}\right) \
        e^{i m \phi_{\vec{R}}}.
    \end{aligned}
    \label{eq:valence_neutron_container}
\end{equation}
Here $m$ is the magnetic quantum number, which is set to be $+1$ and $-1$ for the two valence neutrons to achieve an overall zero angular momentum, and $\phi_{\vec{R}}$ is the azimuthal angle of the neutron position $\vec{R}$ in the spherical coordinate system.

\begin{figure}[ht]
  \centering
  \includegraphics[scale=0.12]{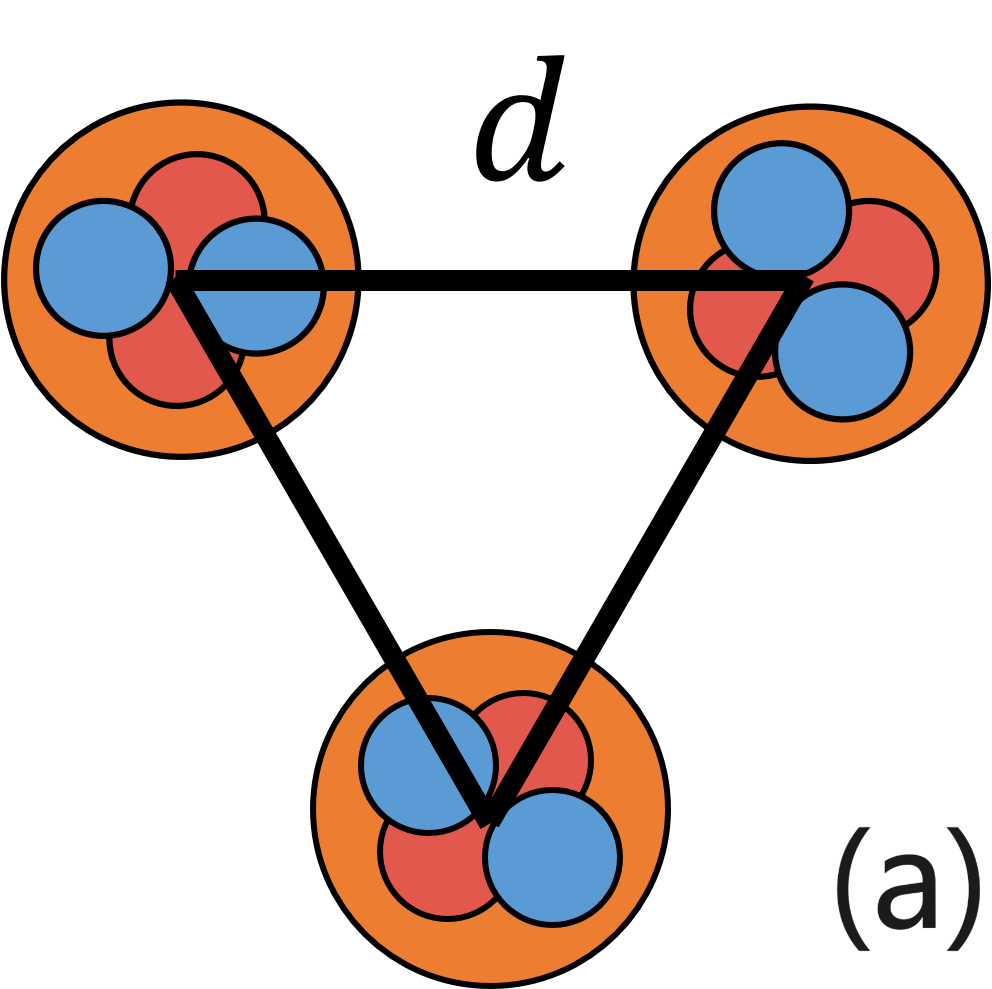}~~~~~ \includegraphics[scale=0.12]{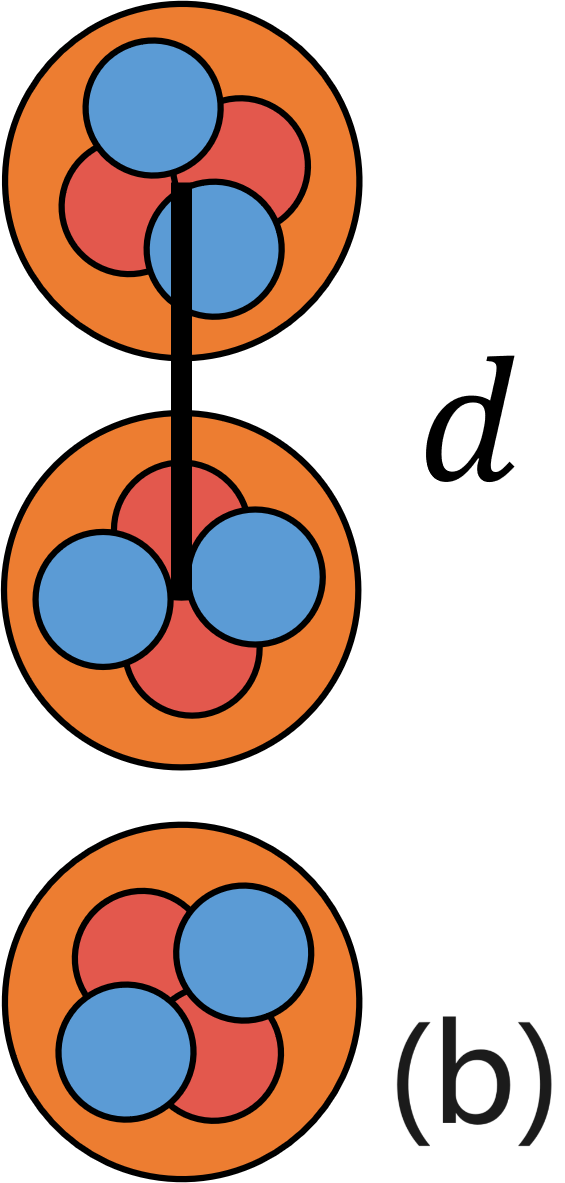}~~~~~
  \includegraphics[scale=0.12]{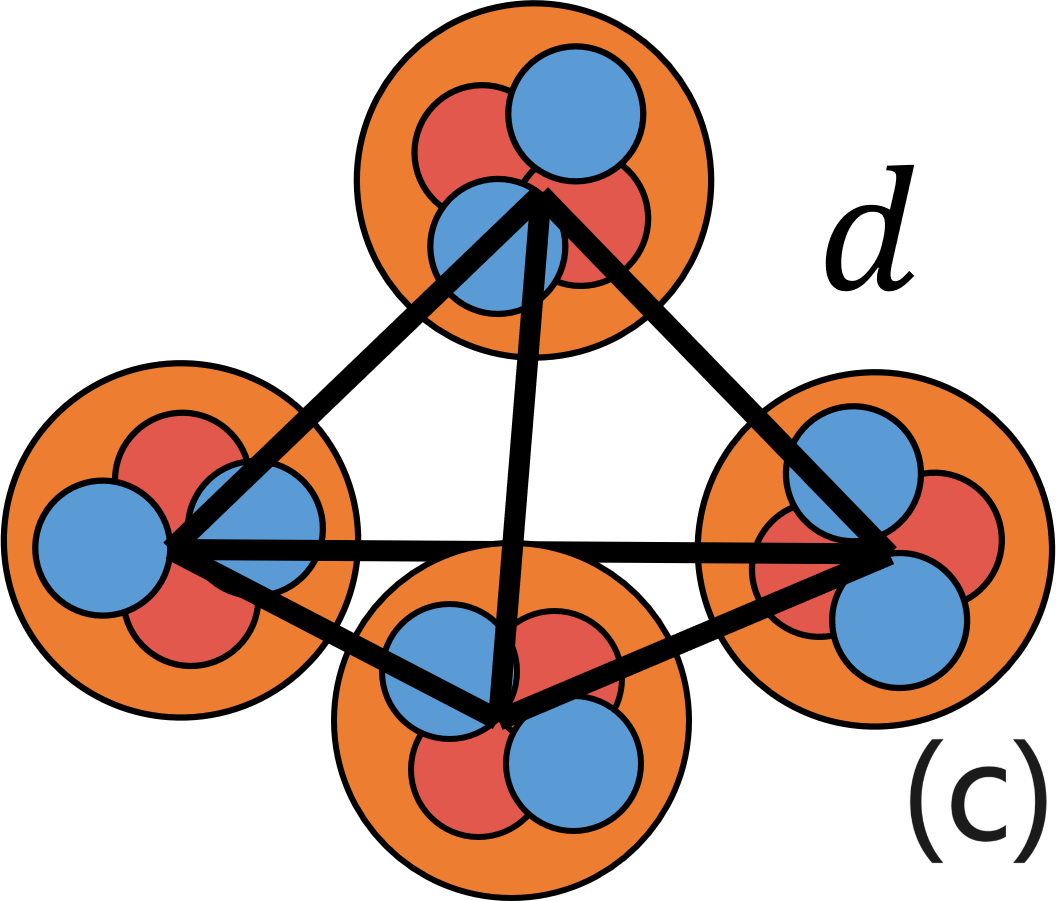}~~~~~ \includegraphics[scale=0.12]{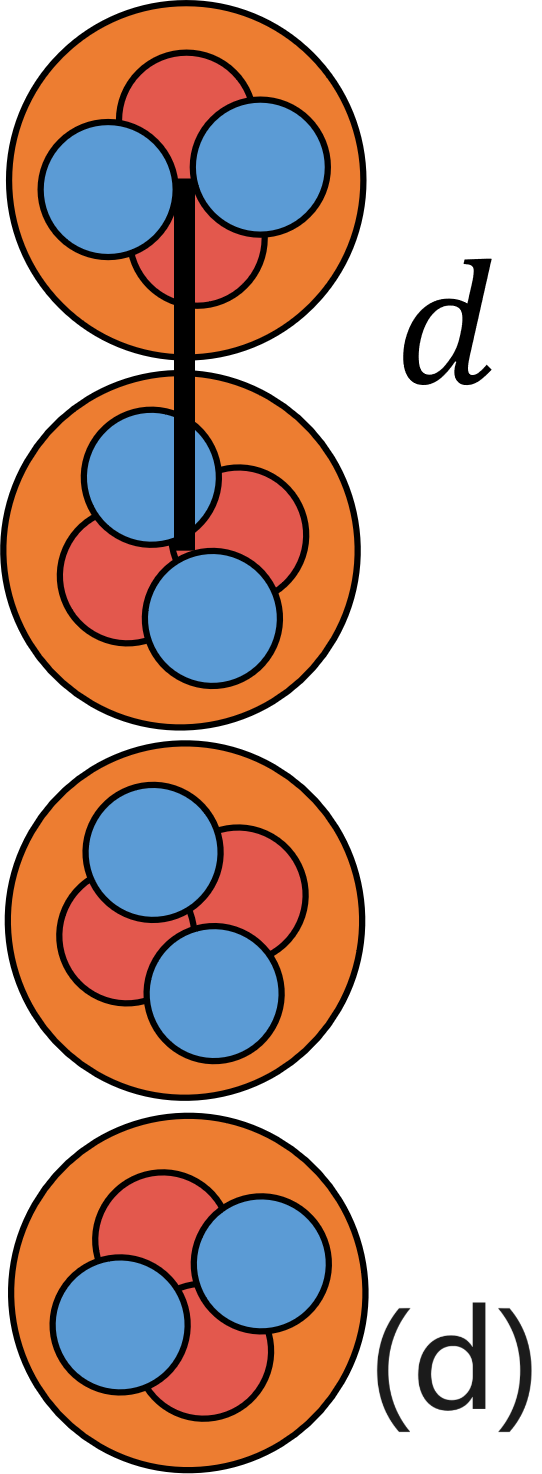} \\
\vspace{1cm}
  \includegraphics[scale=0.12]{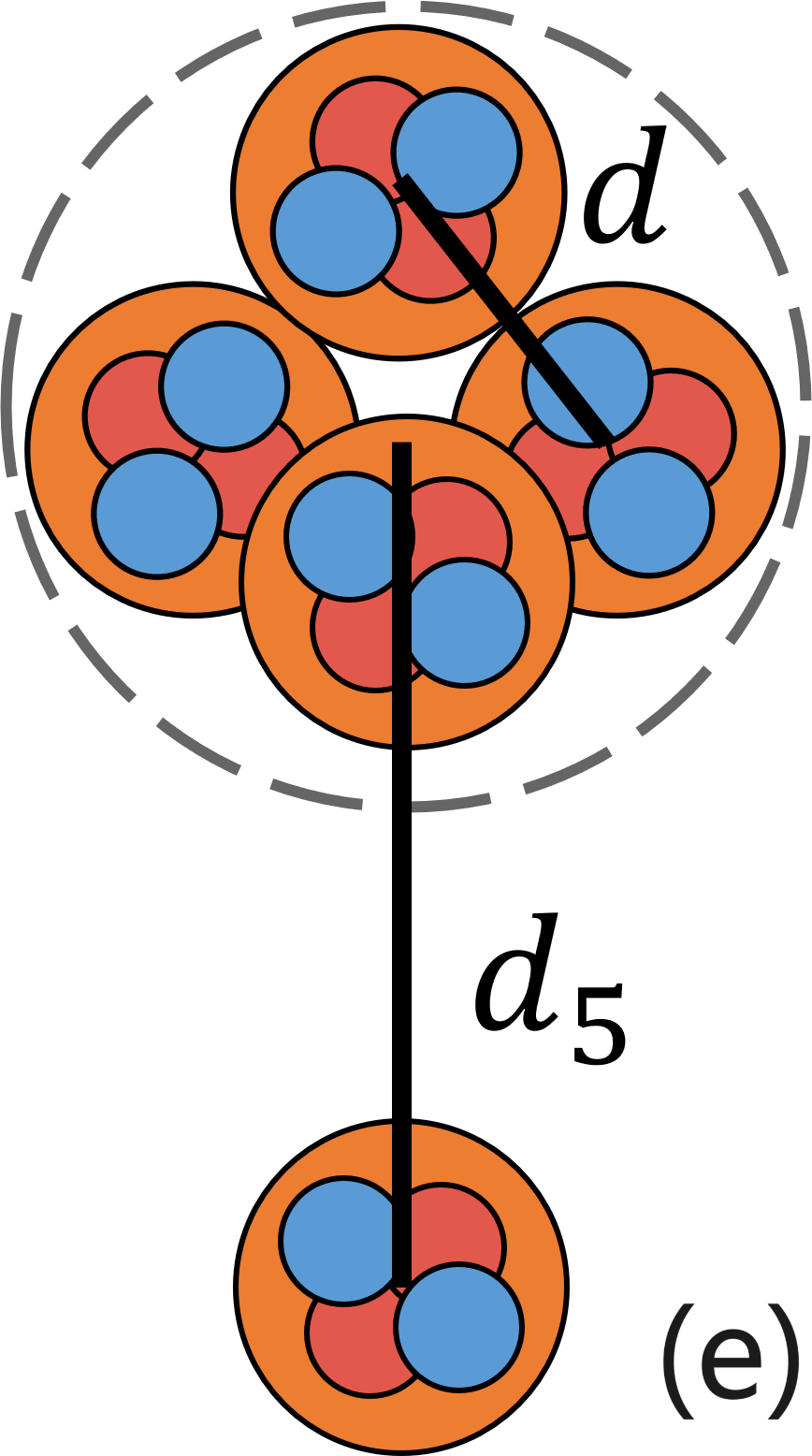}~~~~~~ \includegraphics[scale=0.12]{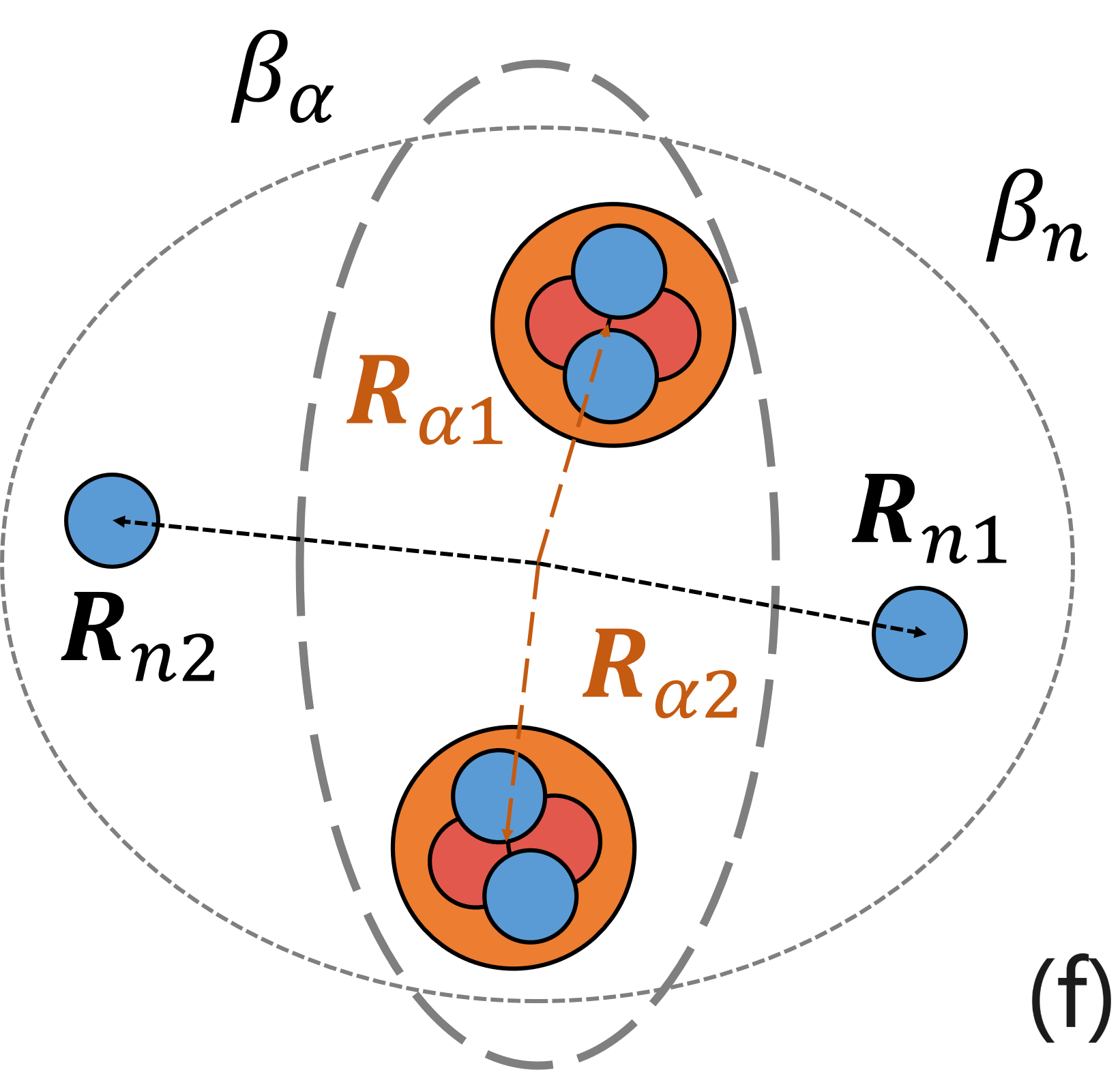}
  \caption{(a) Triangle configuration of $^{12}$C; (b) Chain configuration of $^{12}$C; (c) Tetrahedron configuration of $^{16}$O; (d) Chain configuration of $^{16}$O; (e) 5-$\alpha$ configuration of $^{20}$Ne; (f) Configuration of two $\alpha$ clusters and two valence neutrons for $^{10}$Be.}
  \label{conf}
\end{figure}

The above wave functions should then be further antisymmetrized, and they are calculated in the c.m. frame of the nuclei. To restore the rotational symmetry, we further project the wave functions through
\begin{equation}
  | \Psi^{J}_{M} \rangle
  = \hat{\mathcal{P}}^{J}_{M,K} | \Phi^{\textup{Brink/THSR}} \rangle,
  \label{eq:total_wave_function}
\end{equation}
where the corresponding angular momentum projection operator is expressed as~\cite{Ring_2004_Book}
\begin{eqnarray}
  \hat{\mathcal{P}}^{J}_{M,K} =
  \frac{2J+1}{16\pi^2} \ \int^{2\pi}_0 d\phi \
  \int^{\pi}_0 d \theta\ \mathrm{sin}(\theta) \notag \\
 \times \int^{4\pi}_0 d\gamma \
  D^{J*}_{M,K}(\phi, \theta, \gamma)
  \hat{R}(\phi, \theta, \gamma),
  \label{eq:angular_momentum_projection}
\end{eqnarray}
with $D^{J}_{M,K}(\phi, \theta, \gamma)$ being the Wigner rotation matrix, $\hat{R}$ being the rotation operator, and $\phi$, $\theta$, and $\gamma$ being the three Euler angles. $J$ and $M$ are respectively the quantum number of the total angular momentum and that in the third direction after projection, and $K$ is the one before projection. Since the ground states of $^{12}$C, $^{16}$O, $^{20}$Ne, and $^{10}$Be are all $0^+$ states, both $J$ and $M$ are set to be 0. The wave function $| \Psi^{J}_{M} \rangle$ after the angular momentum projection is used to calculate the energy of the ground state, and the variational principle is used in order to find the minimum energy with the optimized distance among nucleons for specific configurations as shown in Fig.~\ref{conf}. For $^{12}$C, we consider typical triangle and linear-chain configurations, as shown in Fig.~\ref{conf} (a) and Fig.~\ref{conf} (b). For $^{16}$O, we consider typical tetrahedron and linear-chain configurations, as shown in Fig.~\ref{conf} (c) and Fig.~\ref{conf} (d). The distance parameter $d$ in the configurations of $^{12}$C and $^{16}$O is varied in order to get the minimum energy. For $^{20}$Ne, we take the structure of a tetrahedron configuration for four $\alpha$ clusters and another one under the bottom of the tetrahedron~\cite{Yamaguchi:2023mya}, as shown in Fig.~\ref{conf} (e), and the distance parameters $d$ and $d_5$ are varied. For $^{10}$Be, as shown in Fig.~\ref{conf} (f), values of parameters for the Gaussian container are all varied to achieve the ground-state energy.

With the optimized distance for a specific configuration obtained, we use the wave function before the angular momentum projection to calculate the density distribution
\begin{equation}
  \rho(\vec{a})
  = \frac{\langle \Phi | \sum_{i=1}^A \delta(\vec{r}_i - \vec{a}) | \Phi \rangle}
  {\langle \Phi | \Phi \rangle}
  = \sum_{i=1}^A \rho_i(\vec{a}),
\end{equation}
where the density for the $i$th nucleon at position $\vec{a}$ can be calculated from
\begin{eqnarray}
  \rho_i(\vec{a})
  &=& \frac{\langle \Phi | \delta(\vec{r}_i - \vec{a}) | \Phi \rangle}{\langle \Phi | \Phi \rangle} \notag\\
  &=& \frac{1}{(2 \pi)^3} \int d^3 {k} \ e^{-i \vec{k} \cdot \vec{a}}
  \frac{\langle \Phi | e^{i \vec{k} \cdot \vec{r}_i} | \Phi \rangle}
  {\langle \Phi | \Phi \rangle}.
\end{eqnarray}
The initial nucleons in relativistic heavy-ion collisions are sampled according to the above density distributions, and the later dynamics is modeled by the string-melting version of the AMPT model, to be discussed in the next subsection. It is noteworthy that we first sample the nucleons within each $\alpha$ cluster, and then use the corresponding distance parameters to construct the specific configuration, so in this way the $\alpha$-cluster structure is preserved in the sampling.

\subsection{A multiphase transport model}

In the AMPT model~\cite{Lin:2004en}, the initial particle production in relativistic heavy-ion collisions is modelled by a heavy ion jet interaction generator (HIJING) model~\cite{Wang:1991hta}, where the Lund string fragmentation function
\begin{equation}\label{lund}
f(z) \propto z^{-1} (1-z)^a \exp(-b m_\perp^2/z)
\end{equation}
is used to describe the momentum spectrum of the produced particles, with $z$ being the light-cone momentum fraction of the produced hadron of transverse mass $m_\perp$ with respect to that of the fragmenting string, and $a$ and $b$ being two paramters. In the string-melting version, these particles are converted to their valence quarks and antiquarks at the same spatial coordinates. Partons do not undergo scatterings until they have propagated for a given formation time. The later dynamics of these partons is described by Zhang's parton cascade (ZPC) model~\cite{Zhang:1997ej}, where two-body elastic scatterings between partons are simulated using the following differential cross section
\begin{equation}\label{xsection}
\frac{d\sigma}{dt} \approx \frac{9\pi \alpha_s^2}{2(t-\mu^2)^2},
\end{equation}
with $t$ being the standard Mandelstam variable for four-momentum transfer, $\alpha_s$ being the strong coupling constant, and $\mu$ being the screening mass in the partonic matter. After the kinetic freeze-out of these partons, quarks and antiquarks are converted to hadrons via a spatial coalescence model. The later dynamics of the hadronic phase is described by a relativistic transport (ART) model~\cite{Li:1995pra} with various hadronic elastic and inelastic scattering and decay channels.

In the present study, we set the values of the parameters to be $a=0.5$ and $b=0.9$ GeV$^{-2}$ in the Lund string fragmentation function [Eq.~(\ref{lund})], and $\alpha_s=0.33$ and $\mu=3.2$ fm$^{-1}$ in the parton scattering cross section [Eq.~(\ref{xsection})]. These parameters have been shown to reproduce the particle multiplicity and anisotropic flows in Au+Au collisions at $\sqrt{s_{NN}}=200$ GeV~\cite{Xu:2011fe} and Pb+Pb collisions at $\sqrt{s_{NN}}=2.76$ TeV~\cite{Xu:2011fi} reasonably well.

\subsection{Probes for deformation of colliding nuclei}
\label{probes}

In this subsection, we give a brief discussion of the probes for the deformation of colliding nuclei in relativistic heavy-ion collisions. Let's first assume that the nucleus density distribution can be approximately described by an axial symmetric deformed WS form, i.e.,
\begin{equation}\label{dfms}
\rho(r,\theta) = \frac{\rho_0}{1+\exp[(r-R(\theta))/a]}.
\end{equation}
In the above, $\rho_0$ is the normalization constant, $a$ is the diffuseness parameter, and
\begin{equation}\label{WSR}
R(\theta)=R_0[1+\sum_{n} \beta_n Y_{n,0}(\theta)]
\end{equation}
is the deformed radius, with $R_0$ being the average radius, $\beta_n$ being the deformation parameters, and $Y_{n,0}$ being the spherical harmonics.

For a given density distribution with axial symmetry, the deformation parameter $\beta_n^\star$ can be calculated from
\begin{equation} \label{betastar}
\beta_{n}^\star = \frac{4\pi Q_{n}}{3 A R_{rms}^n},
\end{equation}
where $A$ is the total nucleon number, $R_{rms}$ is the root-mean-square (rms) radius, and
\begin{equation}
Q_n = \int \rho(\bm{r}) r^n Y_{n,0}(\theta) d^3 r
\end{equation}
is the intrinsic multipole moment. An axial symmetric density distribution with a $\beta_n$ in Eq.~(\ref{dfms}) generally leads to a different $\beta_n^\star$~\cite{PhysRevLett.130.212302}. The previous studies, e.g., Refs.~\cite{Zhang:2021kxj,Giacalone:2021udy,STAR2024}, mostly extract $\beta_n$ in Eq.~(\ref{WSR}) rather than $\beta_n^\star$ in Eq.~(\ref{betastar}).

In the present study, we focus on the anisotropic flows $\langle v_n^2 \rangle$, transverse momentum fluctuations $\langle \delta p_T^2 \rangle$, and their correlations $\langle v_n^2 \delta p_T \rangle$ as deformation probes in relativistic heavy-ion collisions, where $\langle...\rangle$ represents the event average. The $n$th-order anisotropic flow $v_n$ originates from the $n$th-order anisotropy coefficient $\epsilon_n$ of the overlap region with respect to the event plane $\Phi_n$. For tip-tip relativistic heavy-ion collisions at zero impact parameter, i.e., with symmetric axis head-on, $\epsilon_n$ can be formally expressed as
\begin{equation}\label{epn}
\epsilon_n e^{in\Phi_n}=-\frac {\int r^n \sin^n(\theta) e^{in\phi} \rho(r,\theta)d^3r}{\int r^n \sin^n(\theta)  \rho(r,\theta)d^3r}.
\end{equation}
For other collision configurations, the orientations of the colliding nuclei can be generated by incorporating the Wigner rotation matrix (see Appendix~\ref{app}). $\epsilon_n$ for an arbitrary collision configuration can also be expressed as
\begin{equation}
\epsilon_n e^{in\Phi_n}=-\frac {\int r^n_\perp e^{in\phi} \rho_\perp (r_\perp,\phi) r_\perp dr_\perp d\phi}{\int r^n_\perp \rho_\perp(r_\perp,\phi)r_\perp dr_\perp d\phi},
\end{equation}
where $\rho_\perp(r_\perp, \phi) = \int \rho dz$ is the transverse nucleon density, with $\rho$ being the nucleus density distribution at the given orientation. The transverse momentum fluctuation $\langle \delta p_T^2 \rangle$ originates from the fluctuation $\langle \delta d_\perp^2\rangle$ of the overlap's inverse area $d_\perp = 1/\sqrt{\overline{x^2}~\overline{y^2}}$~\cite{PhysRevC.102.034905}, where $\overline{(...)}$ represents the average value of a quantity in one event. For tip-tip relativistic heavy-ion collisions at zero impact parameter, $\overline{x^2}$ and $\overline{y^2}$ can be formally expressed as
\begin{eqnarray}
\overline{x^2}&=&\frac {\int r^2 \sin^2(\theta) \cos^2(\phi) \rho(r,\theta)d^3r}{\int\rho(r,\theta)d^3r},\label{x2}\\
\overline{y^2}&=&\frac {\int r^2 \sin^2(\theta) \sin^2(\phi) \rho(r,\theta)d^3r}{\int\rho(r,\theta)d^3r},\label{y2}
\end{eqnarray}
respectively. For other collision configurations, again, the orientations of the colliding nuclei can be generated by incorporating the Wigner rotation matrix (see Appendix~\ref{app}). The correlation between the anisotropic flow and the transverse momentum fluctuation $\langle v_n^2 \delta p_T \rangle$ thus originates from that between the anisotropic coefficient and the overlap's inverse area $\langle \epsilon_n^2 \delta d_\perp \rangle$.

Following the procedure in Refs.~\cite{Jia:2021tzt,Jia:2021qyu}, we consider a uniform density distribution with a sharp edge, i.e., $a=0$ in Eq.~(\ref{dfms}). Here we slightly go beyond the derivation in Refs.~\cite{Jia:2021tzt,Jia:2021qyu} by considering more higher-order $\beta_n$ terms in the expansion of the numerator in Eqs.~(\ref{epn}), (\ref{x2}), and (\ref{y2}). With only nonzero $\beta_2$ and $\beta_3$ in Eq.~(\ref{WSR}) as considered in the present study, we can get the following approximate relations
\begin{eqnarray}
\langle \epsilon_2^2 \rangle &=&0.477\beta_2^2+0.172\beta_2^3+0.161\beta_2\beta_3^2+O(\beta_n^4),\label{ep2}\\
\langle \epsilon_3^2 \rangle &=&0.539\beta_3^2+0.452\beta_2^2\beta_3+O(\beta_n^4),\label{ep3}\\
\langle \delta d_\perp^2\rangle &=& \frac{0.25}{R_0^4}\times(7.954\beta_2^2-4.301\beta_2^3+5.352\beta_2\beta_3^2)\notag\\
&&+O(\beta_n^4), \label{dd2} \\
\langle \epsilon_2^2 \delta d_\perp \rangle &=& \frac{0.05}{R_0^2}\times(-8.602\beta_2^3-3.875\beta_2^4-8.682\beta_2^2\beta_3^2)\notag\\
&&+O(\beta_n^5),\label{ep2dd2}\\
\langle \epsilon_3^2 \delta d_\perp \rangle &=& \frac{0.05}{R_0^2}\times(-7.849\beta_2\beta_3^2-0.440\beta_2^2\beta_3^2-3.813\beta_3^4)\notag\\
&&+O(\beta_n^5). \label{ep3dd2}
\end{eqnarray}

If $\langle v_n^2 \rangle$, $\langle \delta p_T^2 \rangle$, and $\langle v_n^2 \delta p_T \rangle$ are linearly correlated with $\langle \epsilon_n^2 \rangle$, $\langle \delta d_\perp^2\rangle$, and $\langle \epsilon_n^2 \delta d_\perp \rangle$, respectively, similar relations between $\langle v_n^2 \rangle$, $\langle \delta p_T^2 \rangle$, and $\langle v_n^2 \delta p_T \rangle$ and $\beta_n$ are expected to be valid. For the detailed derivation to get Eqs.~(\ref{ep2})-(\ref{ep3dd2}), we refer the reader to Appendix~\ref{app}. In the lowest order, these relations are similar to those in Refs.~\cite{Jia:2021tzt,Jia:2021qyu}, while higher-order $\beta_n$ terms as well as cross terms appear when we consider large deformation for light nuclei. These relations are useful for the illustration purpose, while one should keep in mind that there are a few approximations and limitations of these relations. First, we neglected event-by-event fluctuations and used uniform density distributions with sharp surfaces. Here the event-by-event fluctuations are caused by finite particle numbers and stochastic dynamics. Second, the derivation is only valid for not too large $\beta_n$ but may fail for $\beta_n \sim 1$.

\section{results and discussions}
\label{results}

The purpose of the present study is to investigate whether the deformation probes work for light nuclei with both large $\beta_n$ and special internal structures. In this section, we will first evaluate the performance of the deformation probes for nuclei with large $\beta_2$ or $\beta_3$, and then discuss how these deformation probes work for light nuclei with different $\alpha$-cluster configurations.

\subsection{Validity of probes at large deformation}
\label{Zr96}

\begin{figure*}[ht]
\centering
  \includegraphics[scale=0.5]{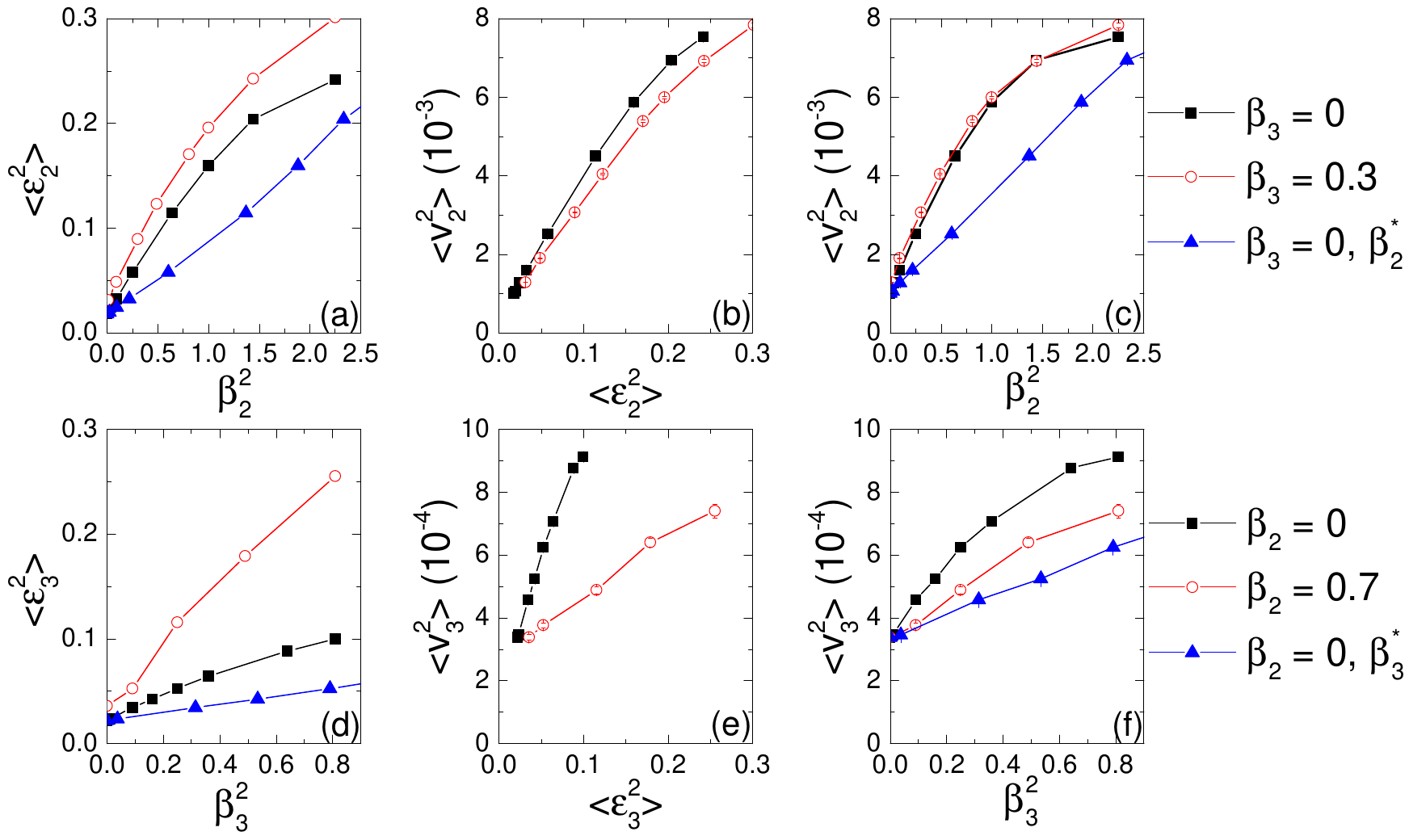}
  \caption{Relations between $\beta^2_{2(3)}$ (${\beta^\star_{2(3)}}^2$), $\langle \epsilon_{2(3)}^2 \rangle$, and $\langle v_{2(3)}^2 \rangle$ for a fixed $\beta_{3(2)}$ in central ($0-5\%$) $^{96}$Zr+$^{96}$Zr collisions at $\sqrt{s_{NN}}=200$ GeV from AMPT calculations.}
  \label{bv23}
\end{figure*}

\begin{figure}[ht]
\centering
  \includegraphics[scale=0.5]{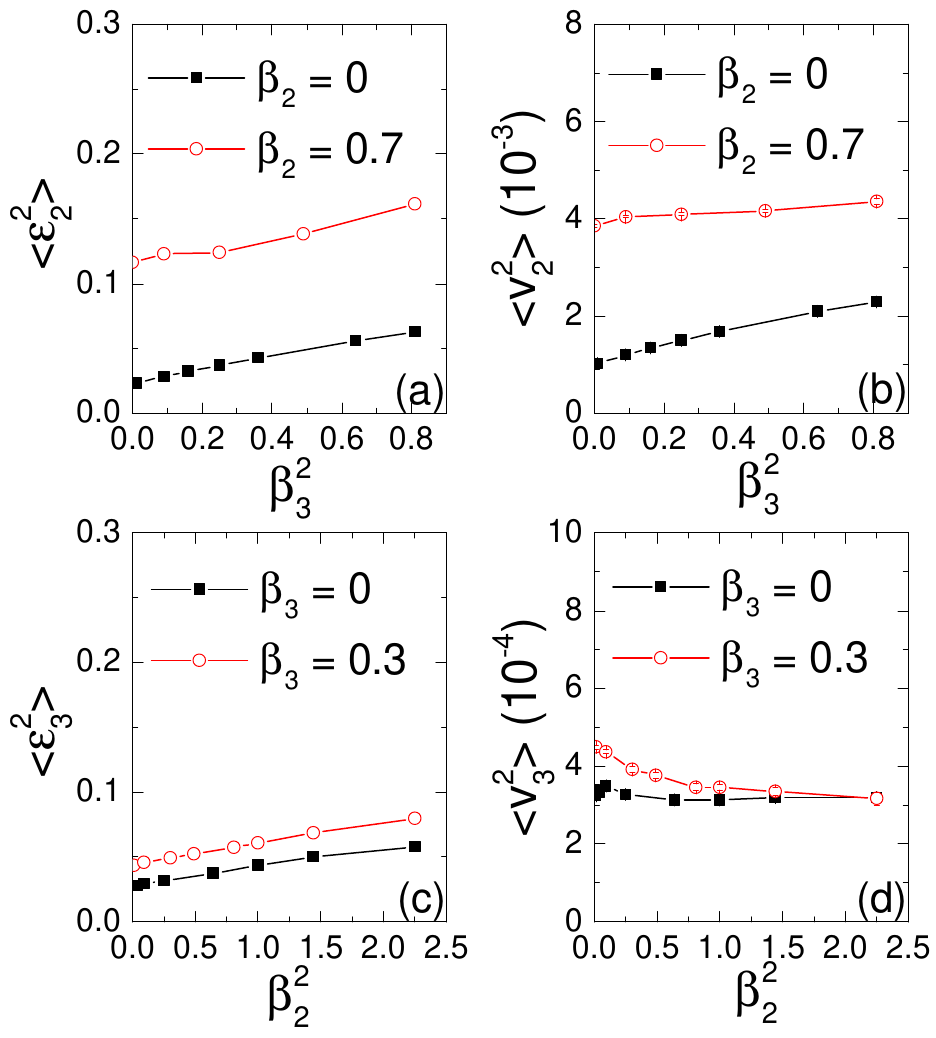}
  \caption{Cross relation between $\beta^2_{3(2)}$ and $\langle \epsilon_{2(3)}^2 \rangle$ and that between $\beta^2_{3(2)}$ and $\langle v_{2(3)}^2 \rangle$ for a fixed $\beta_{2(3)}$ in central ($0-5\%$) $^{96}$Zr+$^{96}$Zr collisions at $\sqrt{s_{NN}}=200$ GeV from AMPT calculations.}
  \label{bv23cross}
\end{figure}

\begin{figure*}[ht]
\centering
  \includegraphics[scale=0.5]{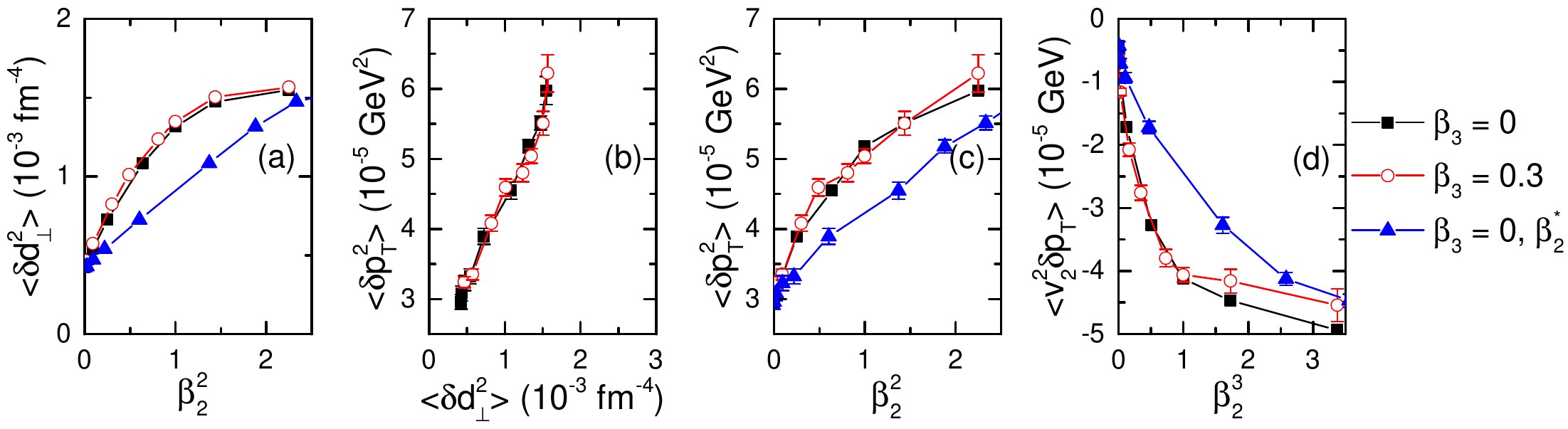}
  \includegraphics[scale=0.5]{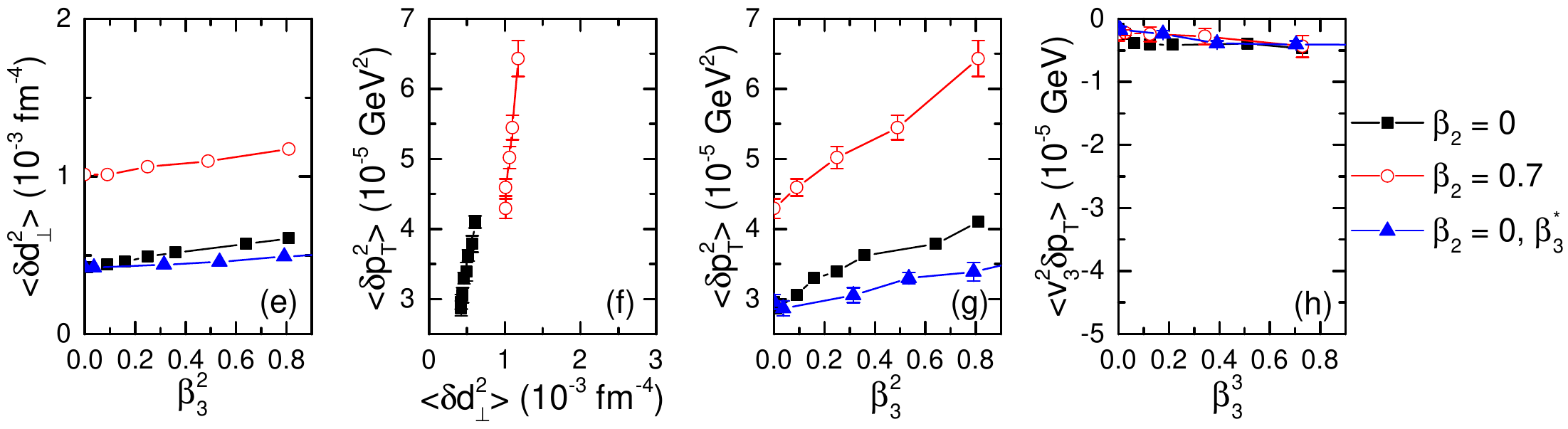}
  \caption{Relations between $\beta^2_{2(3)}$ (${\beta^\star_{2(3)}}^2$), $\langle \delta d_\perp^2 \rangle$, and $\langle \delta p_T^2 \rangle$ as well as that between $\beta^3_{2(3)}$ (${\beta^\star_{2(3)}}^3$) and $\langle v_{2(3)}^2 \delta p_T \rangle$ for a fixed $\beta_{3(2)}$ in central ($0-5\%$) $^{96}$Zr+$^{96}$Zr collisions at $\sqrt{s_{NN}}=200$ GeV from AMPT calculations.}
  \label{b23pt}
\end{figure*}

The performance of the deformation probes is evaluated with $^{96}$Zr+$^{96}$Zr collisions at $\sqrt{s_{NN}}=200$ GeV, and the density distribution of $^{96}$Zr is parameterized as Eqs.~(\ref{dfms}) and (\ref{WSR}) with only nonzero $\beta_2$ and $\beta_3$. We fix $R_0=5.02$ fm and $a=0.52$ fm as in Ref.~\cite{Zhang:2022fou} while vary $\beta_2$ from 0 to 1.5 and $\beta_3$ from 0 to 0.9 to get different deformations of the colliding nuclei, and the dynamics of their collisions is described by the AMPT model. We select events at $0-5\%$ centralities according to charged-particle multiplicities, and investigate the relations between $\langle \epsilon_n^2 \rangle$, $\langle v_n^2 \rangle$, $\langle \delta d_\perp^2 \rangle$, $\langle \delta p_T^2 \rangle$, and $\langle  v_n^2 \delta p_T \rangle$ and the deformation parameters $\beta_n$ in the large ranges of $\beta_2$ and $\beta_3$. The $n$th-order anisotropic coefficient and the fluctuation of the overlap's inverse area are calculated from the coordinates of partons at $t=0$ in AMPT according to\

\begin{equation}
\epsilon_n  =  \frac{ \sqrt{[\sum_i r_{\perp,i}^n \cos(n\phi_i)]^2+[\sum_i r_{\perp,i}^n \sin(n\phi_i)]^2}}{\sum_i r_{\perp,i}^n},
\end{equation}

\vspace{-0.5cm}
\begin{equation}
\delta d_\perp^2 = (d_\perp - \langle d_\perp \rangle)^2,
\end{equation}
respectively, where $r_{\perp,i}=\sqrt{x_i^2+y_i^2}$ and $\phi_i=\arctan(y_i/x_i)$ are respectively the polar coordinate and polar angle of the $i$th particle in the transverse plane, and $d_\perp = 1/\sqrt{\overline{x^2}~\overline{y^2}}$ is the overlap's inverse area with $\overline{(...)}$ representing the average over all particles in one event. $\delta d_\perp$ is linearly correlated with the deviation of the mean transverse momentum away from its event-averaged value $\delta p_T = \overline{p_{T}}- \langle \overline{p_T} \rangle$~\cite{PhysRevC.102.034905}. The anisotropic flows, the transverse-momentum fluctuation, and their correlation are calculated from the phase-space information of particles at the final stage in AMPT according to
\begin{eqnarray}
\langle v_n^2 \rangle &=& \langle \cos [n(\varphi_i-\varphi_j)] \rangle_{i,j}, \\
\langle \delta p_T^2 \rangle &=& \langle (p_{T,i} - \langle \overline{p_T}\rangle) (p_{T,j} - \langle \overline{p_T}\rangle) \rangle_{i,j}, \\
\langle v_n^2 \delta p_T \rangle &=& \langle \cos[n(\varphi_i-\varphi_j)] (p_{T,k} - \langle \overline{p_T}\rangle)\rangle_{i,j,k},
\end{eqnarray}
respectively. Here $\langle...\rangle_{i,j,...}$ represents the average over all possible combinations of $i,j,...$ for all events, and $p_{T,i}=\sqrt{p_{x,i}^2+p_{y,i}^2}$ and $\varphi_i=\arctan(p_{y,i}/p_{x,i})$ are, respectively, the momentum and its polar angle of the $i$th particle in the transverse plane. Particles at midpseudorapidities ($|\eta|<2$) and $0.2<p_T<3$ GeV are selected for the calculation, with a pseudorapidity gap of $|\Delta \eta|>0.5$ used in calculating $\langle...\rangle_{i,j,...}$ to remove the non-flow effect.

Figure~\ref{bv23} displays the relations between $\beta^2_n$, $\langle \epsilon_n^2 \rangle$, and $\langle v_n^2 \rangle$ in central $^{96}$Zr+$^{96}$Zr collisions from AMPT. As shown in the first column, a large $\beta_n^2$ leads to a large $\langle \epsilon_n^2 \rangle$ for both $n=2$ and 3, and the linear relation holds until $\beta_2^2 \gtrsim 1$ and $\beta_3^2 \gtrsim 0.4$. The slower increasing trend at large values of $\beta_{2}^2$ is inconsistent with the positive $\beta_2^3$ term in Eq.~(\ref{ep2}), likely due to the fact that the derivation is not valid at $\beta_2 \gtrsim 1$. On the other hand, the non-zero value of $\langle \epsilon_{2(3)}^2 \rangle$ at $\beta_{2(3)}^2=0$ is due to the effect of finite particle numbers which leads to event-by-event fluctuations. The traditional linear relation between $\langle \epsilon_n^2 \rangle$ and $\langle v_n^2 \rangle$ is shown in the second column, which is seen to be slightly violated at $\langle \epsilon_2^2 \rangle \gtrsim 0.15$ and $\langle \epsilon_3^2 \rangle \gtrsim 0.08$. Consequently, the linear relation between $\beta^2_n$ and $\langle v_n^2 \rangle$ is violated for $\beta_2^2 \gtrsim 1$ and $\beta_3^2 \gtrsim 0.4$. For a finite $\beta_{3(2)}$, $\langle \epsilon_{2(3)}^2 \rangle$ is larger for a given $\beta_{2(3)}^2$, consistent with the positive $\beta_2\beta_3^2$ term in Eq.~(\ref{ep2}) and the positive $\beta_2^2\beta_3$ term in Eq.~(\ref{ep3}), while this generally does not lead to larger anisotropic flows $\langle v_{2(3)}^2 \rangle$. Considering the largest $\beta_2$ and $\beta_3$ for light nuclei obtained in the present study (see Table~\ref{T1}), the linear relation between $\beta^2_n$ and $\langle v_n^2 \rangle$ is mostly valid, as long as the density distribution can be approximately described by a deformed WS form. We have also compared the relations between $\langle \epsilon_n^2 \rangle$, $\langle v_n^2 \rangle$, and ${\beta^\star_n}^2$, with $\beta_n^\star$ calculated from Eq.~(\ref{betastar}). It is interesting to see that the linear relations of $\langle \epsilon_{2(3)}^2 \rangle \sim {\beta^\star_{2(3)}}^2$ and $\langle v_{2(3)}^2 \rangle \sim {\beta^\star_{2(3)}}^2$ are preserved at extremely large $\beta^\star_{2(3)}$, although they have smaller slopes.

Considering that some nuclei (e.g., $^{20}$Ne) have both large $\beta_2$ and $\beta_3$, we discuss more explicitly the cross relation between $\beta^2_{3(2)}$ and $\langle \epsilon_{2(3)}^2 \rangle$ as well as $\langle v_{2(3)}^2 \rangle$ in Fig.~\ref{bv23cross}. It is seen that $\langle \epsilon_{2(3)}^2 \rangle$ increases also approximately linearly with increasing $\beta_{3(2)}$, and for a finite $\beta_{2(3)}$ the whole curves move to the upper side. This is again consistent with the positive $\beta_2\beta_3^2$ term in Eq.~(\ref{ep2}) and the positive $\beta_2^2\beta_3$ term in Eq.~(\ref{ep3}). For the final anisotropic flows, the relation between $\langle v_{2}^2 \rangle$ and $\beta_3^2$ is similar to that between $\langle \epsilon_{2}^2 \rangle$ and $\beta_3^2$, while the relation between $\langle v_{3}^2 \rangle$ and $\beta_2^2$ shows a different behavior. Note that we use the same scale for the y axis in Fig.~\ref{bv23cross} as in Fig.~\ref{bv23}, so it is seen that the correlation strength shown in Fig.~\ref{bv23cross} is much weaker than that shown in Fig.~\ref{bv23}.

Figure~\ref{b23pt} illustrates the relation between $\beta^2_n$, $\langle \delta d_\perp^2 \rangle$, and $\langle \delta p_T^2 \rangle$ as well as between $\beta^3_n$ and $\langle v_n^2 \delta p_T \rangle$ in central $^{96}$Zr+$^{96}$Zr collisions from AMPT calculations. A large $\beta^2_2$ leads to large fluctuations of the initial overlap area characterized by $\langle \delta d_\perp^2 \rangle$, and the linear relation approximately holds until $\beta_2^2 \gtrsim 1$ and becomes saturated, as shown in Fig.~\ref{b23pt} (a). The saturation behavior is consistent with the negative coefficient of the $\beta_2^3$ term in Eq.~(\ref{dd2}). The linear relation between $\langle \delta d_\perp^2 \rangle$ and $\langle \delta p_T^2 \rangle$ is also approximately valid from the AMPT dynamics, as shown in Fig.~\ref{b23pt} (b). Consequently, the linear relation between $\beta^2_2$ and $\langle \delta p_T^2 \rangle$ is approximately valid until $\beta_2^2 \gtrsim 1$ and then the increasing trend becomes slower, as shown in Fig.~\ref{b23pt} (c). We also illustrate the relation between $\beta^3_2$ and $\langle v_2^2 \delta p_T \rangle$ in Fig.~\ref{b23pt} (d). The negative value of $\langle v_2^2 \delta p_T \rangle$ is due to the fact that a larger (smaller) overlap area generally leads to a smaller (larger) $\overline{p_T}$ but a larger (smaller) $v_2^2$, and this is especially so for a larger $\beta_2$ (see typical cases for central tip-tip and body-body collisions). The linear relation between $\beta^3_2$ and $\langle v_2^2 \delta p_T \rangle$ is valid until $\beta_2^3 \gtrsim 0.5$ and then the slope becomes much smaller for larger $\beta_2$. This behavior is inconsistent with the negative $\beta_2^4$ term in Eq.~(\ref{ep2dd2}), probably due to the event-by-event fluctuation or that the derivation is not applicable at too large $\beta_2$. While there are some linear relations between $\beta^2_3$ and $\langle \delta d_\perp^2 \rangle$ as well as $\langle \delta p_T^2 \rangle$ and between $\beta^3_3$ and $\langle v_3^2 \delta p_T \rangle$, the correlation strength is rather weak compared to those for $\beta_2$, as can also be expected from Eq.~(\ref{ep3dd2}) where there is no $\beta_3^3$ term. These relations with a fixed finite $\beta_3$ or $\beta_2$ are also compared, and the qualitative behaviors are consistent with the above discussions. Again, the linear relations between $\beta^2_n$, $\langle \delta d_\perp^2 \rangle$, and $\langle \delta p_T^2 \rangle$ as well as those between $\beta^3_n$ and $\langle v_n^2 \delta p_T \rangle$ are mostly valid for reasonable values of $\beta_2$ and $\beta_3$ in realistic nuclei whose density distributions can be approximately described by a deformed WS form. The linear relations of $\langle \delta d_\perp^2 \rangle \sim {\beta^\star_{2(3)}}^2$, $\langle \delta p_T^2 \rangle \sim {\beta^\star_{2(3)}}^2$, and $\langle v_{2(3)}^2 \delta p_T\rangle \sim {\beta^\star_{2(3)}}^3$ are better preserved, and they have mostly smaller slopes.

\subsection{Validity of probes with $\alpha$ clusters}

\begin{figure*}[ht]
  \centering
  \includegraphics[scale=0.15]{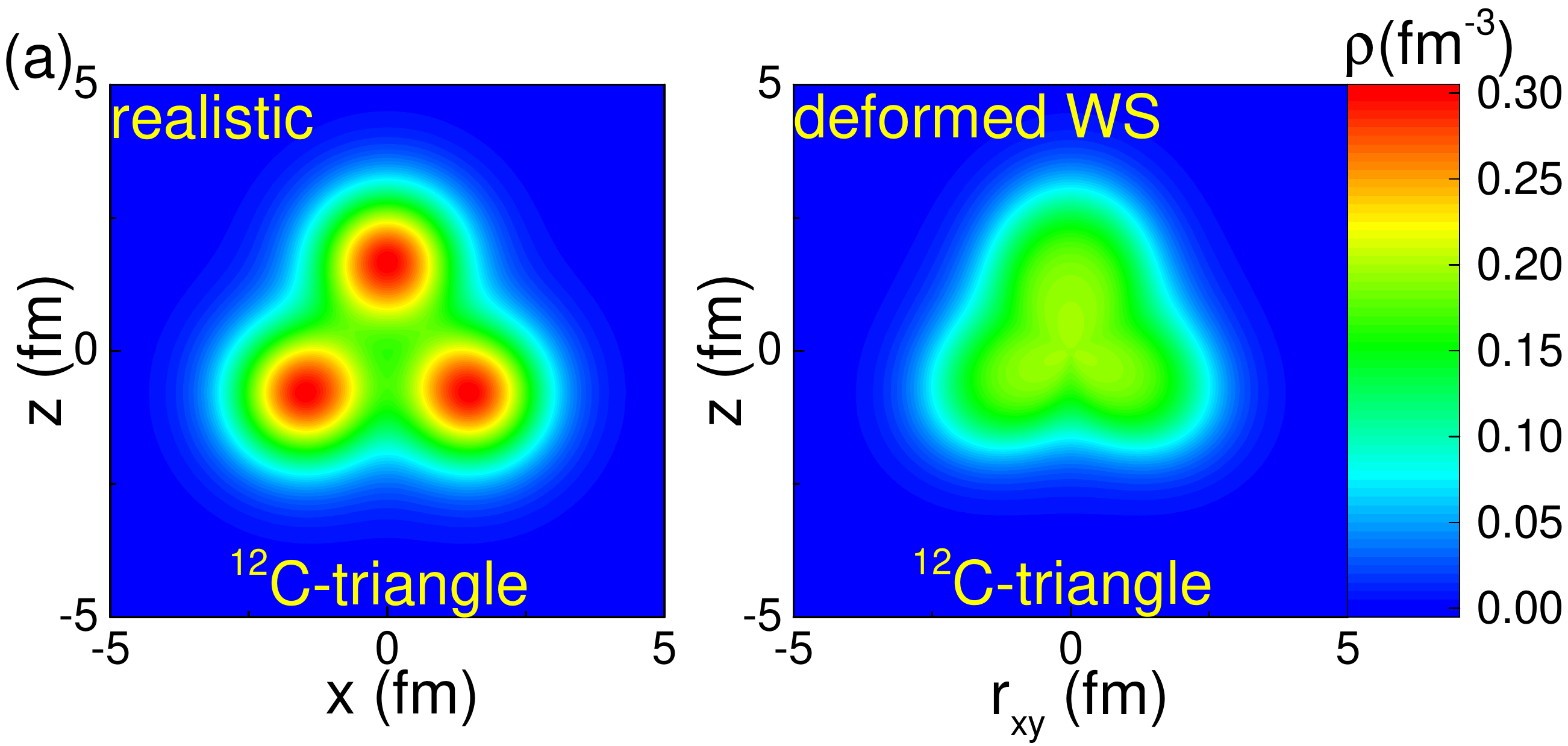}
  \includegraphics[scale=0.15]{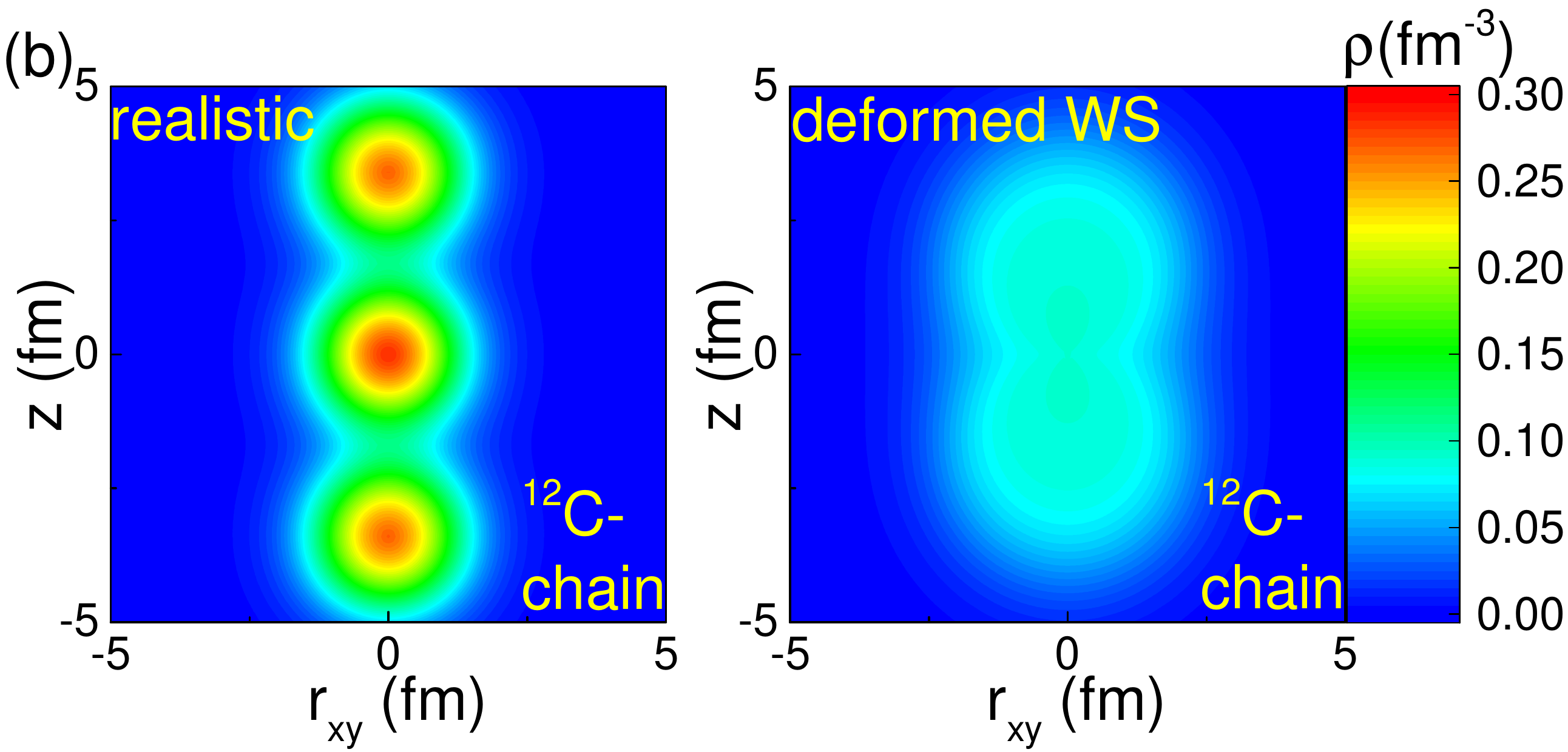}\\
  \includegraphics[scale=0.15]{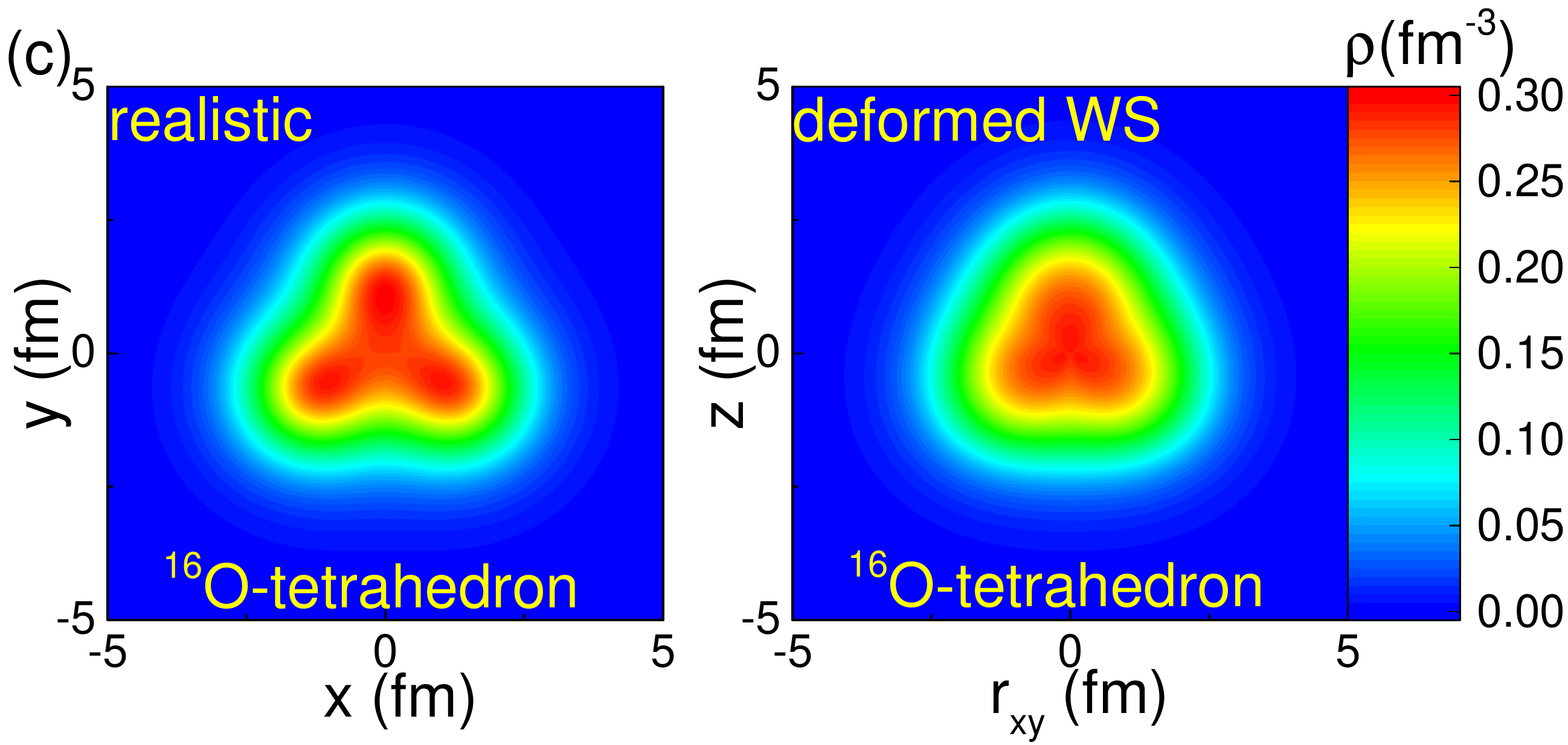}
  \includegraphics[scale=0.15]{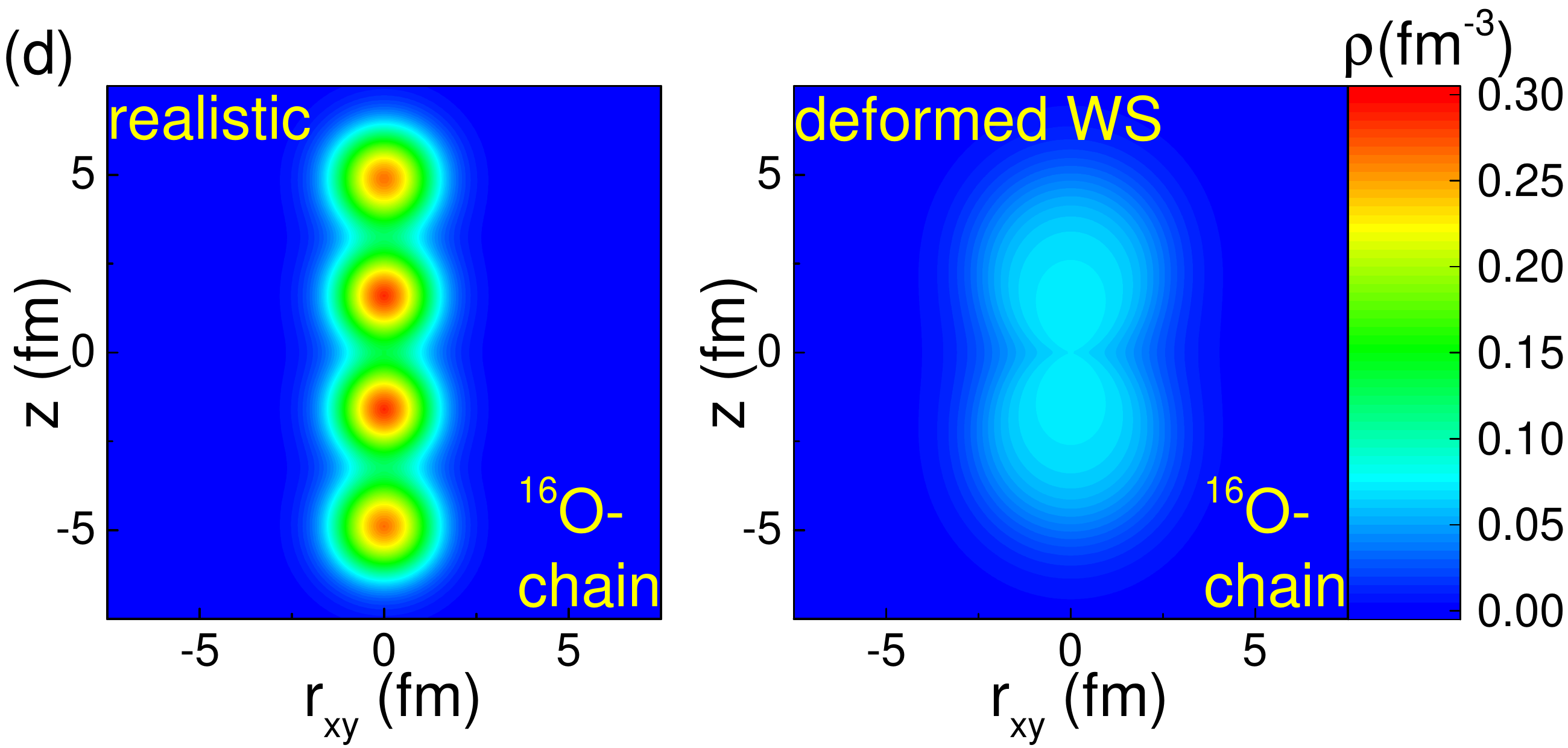}\\
  \includegraphics[scale=0.15]{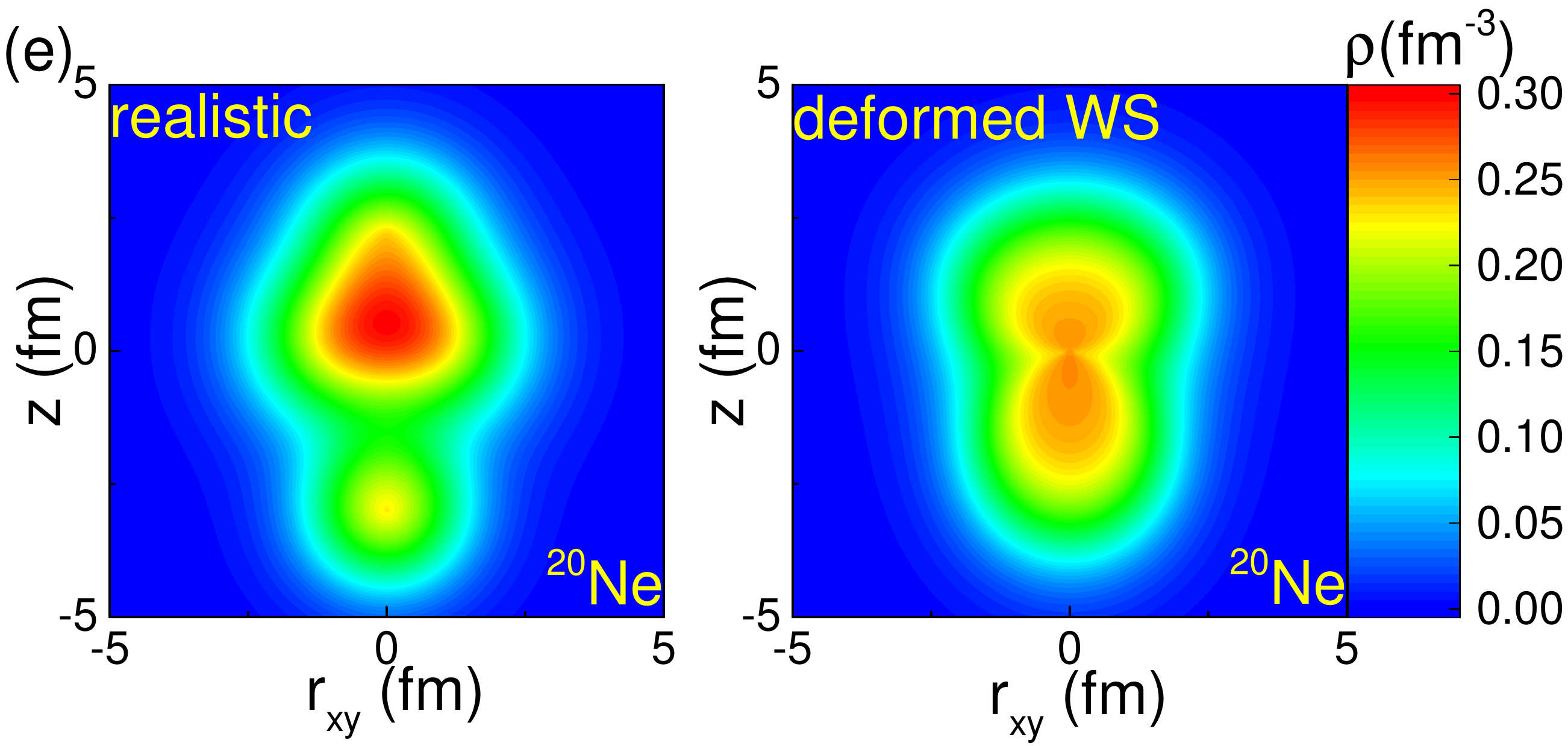}
  \includegraphics[scale=0.15]{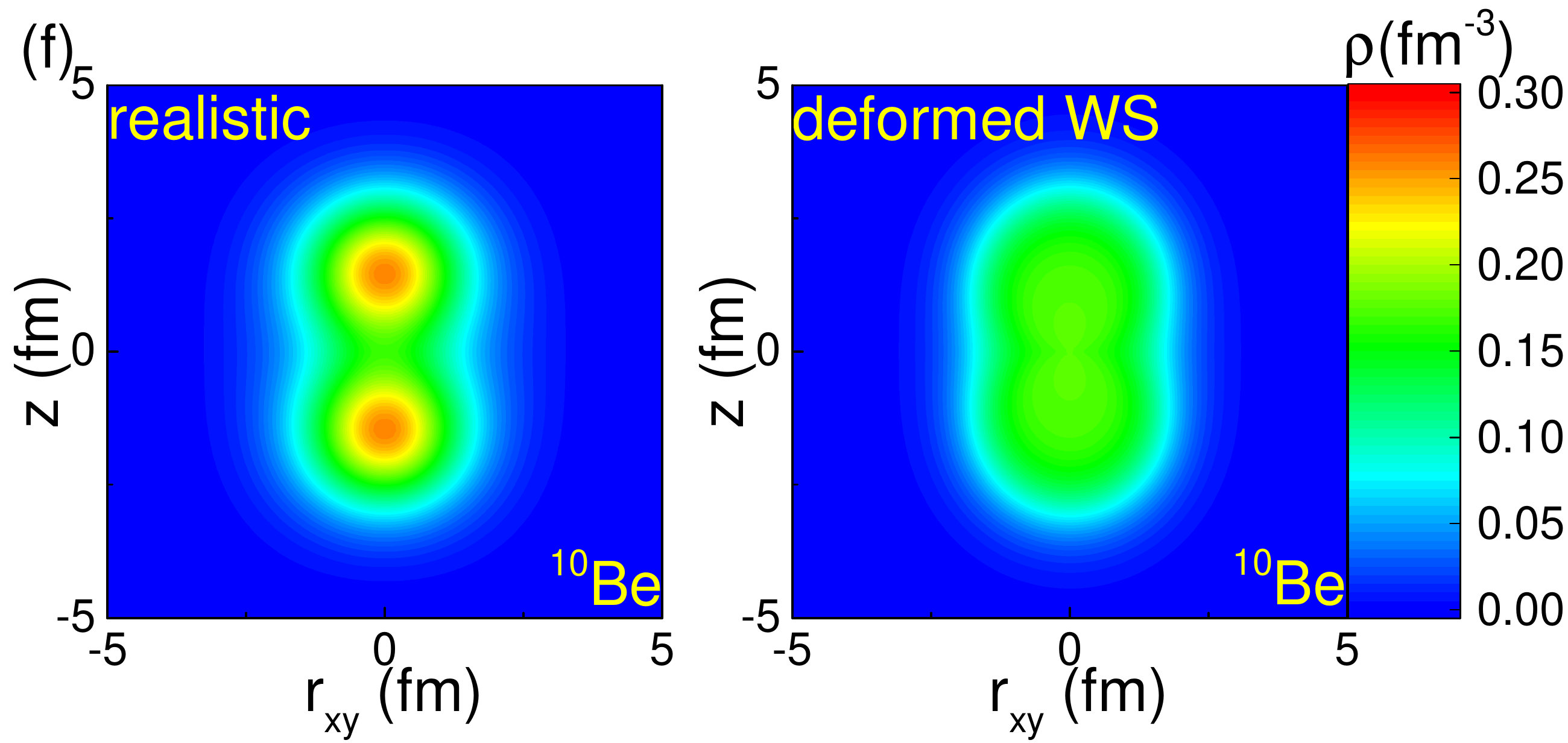}\\
  \caption{Density distributions of triangle (a) and chain (b) configurations of $^{12}$C, tetrahedron (c) and chain (d) configurations of $^{16}$O, $^{20}$Ne (e), and $^{10}$Be (f), from realistic calculations with $\alpha$ clusters and from a deformed WS form with the same $\beta_2^\star$ and $\beta_3^\star$.}
  \label{rho}
\end{figure*}

\begin{table}[h!]
\centering
\caption{Values of $\beta_2^\star$ and $\beta_3^\star$ for different configurations of $^{12}$C, $^{16}$O, $^{20}$Ne, and $^{10}$Be as well as the corresponding values of $\beta_2$ and $\beta_3$ used in the parameterized deformed WS distributions as shown in Fig.~\ref{rho}.}
\label{T1}
\renewcommand\arraystretch{1.5}
\setlength{\tabcolsep}{3mm}
\begin{tabular}{cccccc}
\hline\hline
 & $\beta_2^\star$ & $\beta_3^\star$ & $\beta_2$ & $\beta_3$ \\
\hline
$^{12}$C triangle & 0 & 0.648 & 0 & 0.439 \\
$^{12}$C chain & 0.938 & 0 & 0.783 & 0 \\
$^{16}$O tetrahedron & 0 & 0.300 & 0 & 0.223 \\
$^{12}$O chain & 1.075 & 0 & 1.014 & 0 \\
$^{20}$Ne & 0.710 & 0.448 & 0.666 & 0.250 \\
$^{10}$Be & 0.854 & 0 & 0.693 & 0 \\
\hline\hline
\end{tabular}
\end{table}

\begin{figure*}[ht]
  \centering
  \includegraphics[scale=0.5]{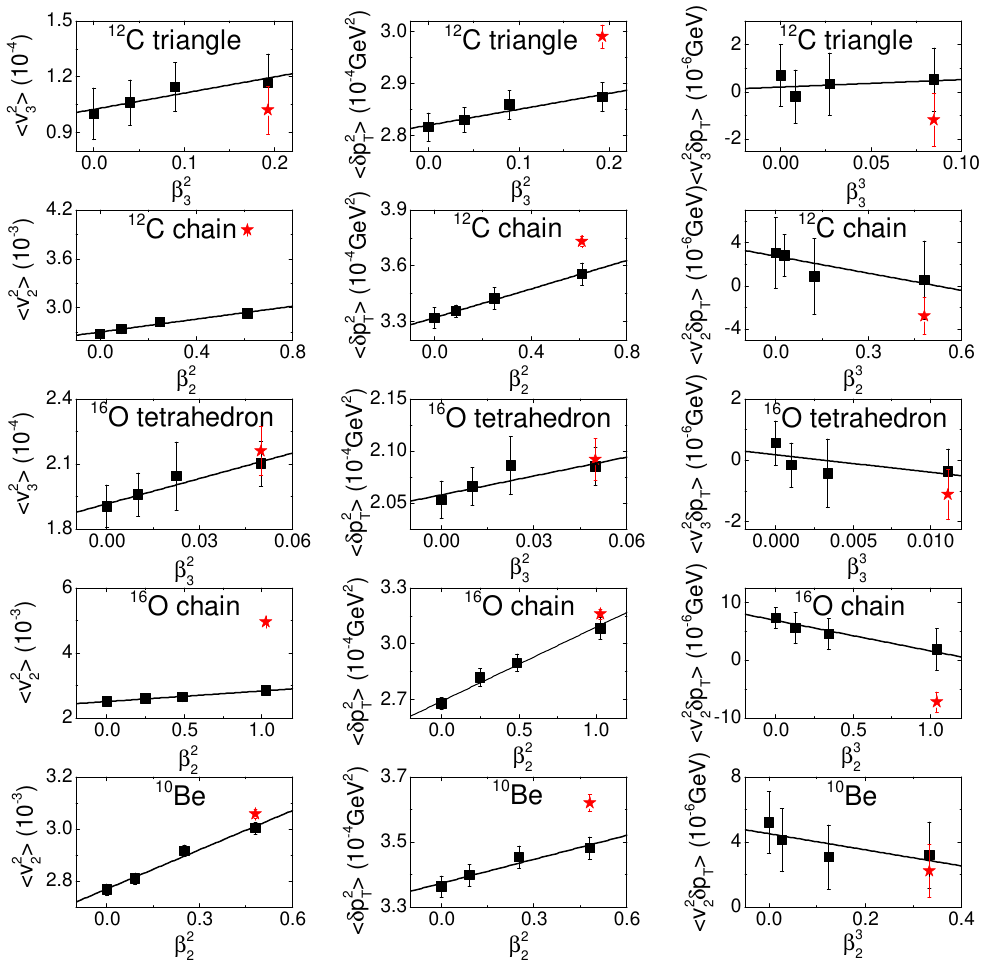}\includegraphics[scale=0.5]{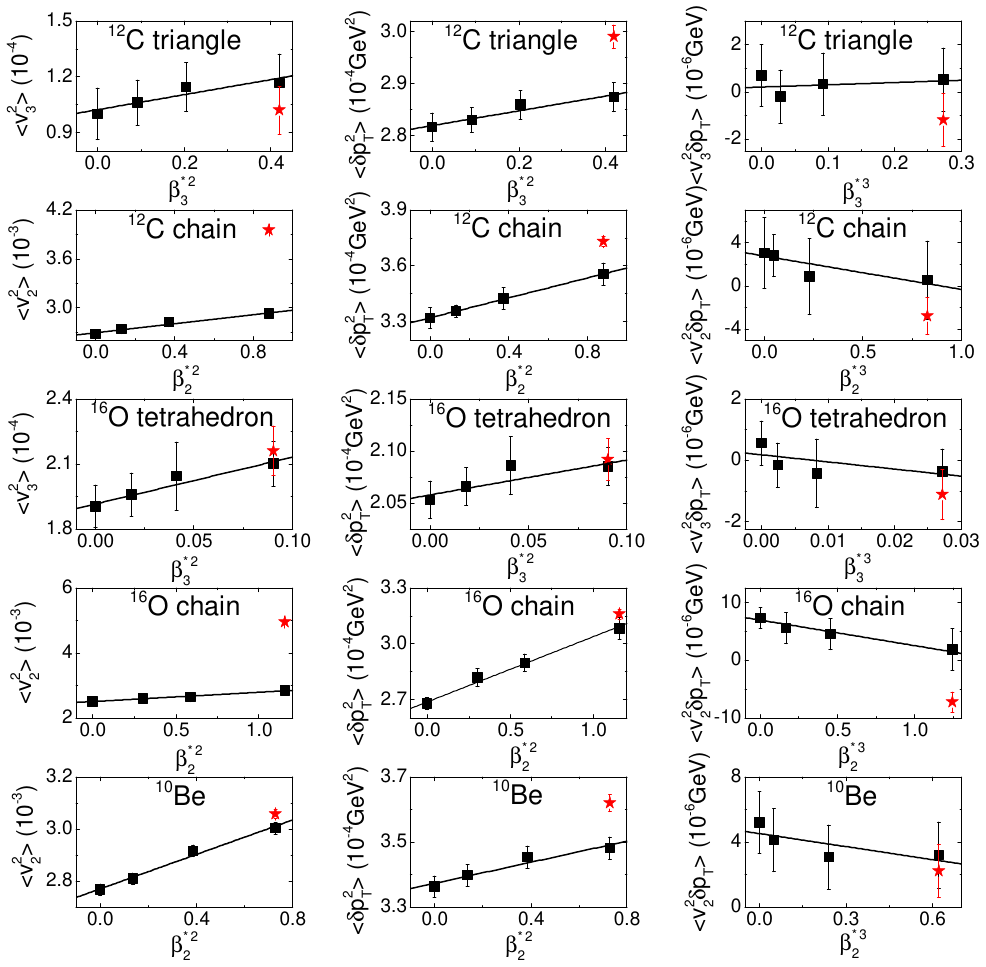}
  \caption{Relations between $\langle v_{2(3)}^2 \rangle$ and $\beta_{2(3)}^2$ (first column), $\langle \delta p_T^2 \rangle$ and $\beta_{2(3)}^2$ (second column), as well as $\langle v_{2(3)}^2\delta p_T \rangle$ and $\beta_{2(3)}^3$ (third column) for triangle (first row) and chain (second row) configurations of $^{12}$C, tetrahedron (third row) and chain (fourth row) configurations of $^{16}$O, and $^{10}$Be (fifth row) in central ($0-5\%$) collisions of these nuclei at $\sqrt{s_{NN}}=200$ GeV. Black squares represent results from density distributions of the WS form, while red stars represent results from realistic density distributions with $\alpha$ clusters. The fourth, fifth, and sixth columns show similar relations but with $\beta^\star_{2(3)}$.}
  \label{lightnuclei}
\end{figure*}

\begin{figure*}[ht]
  \centering
  \includegraphics[scale=0.45]{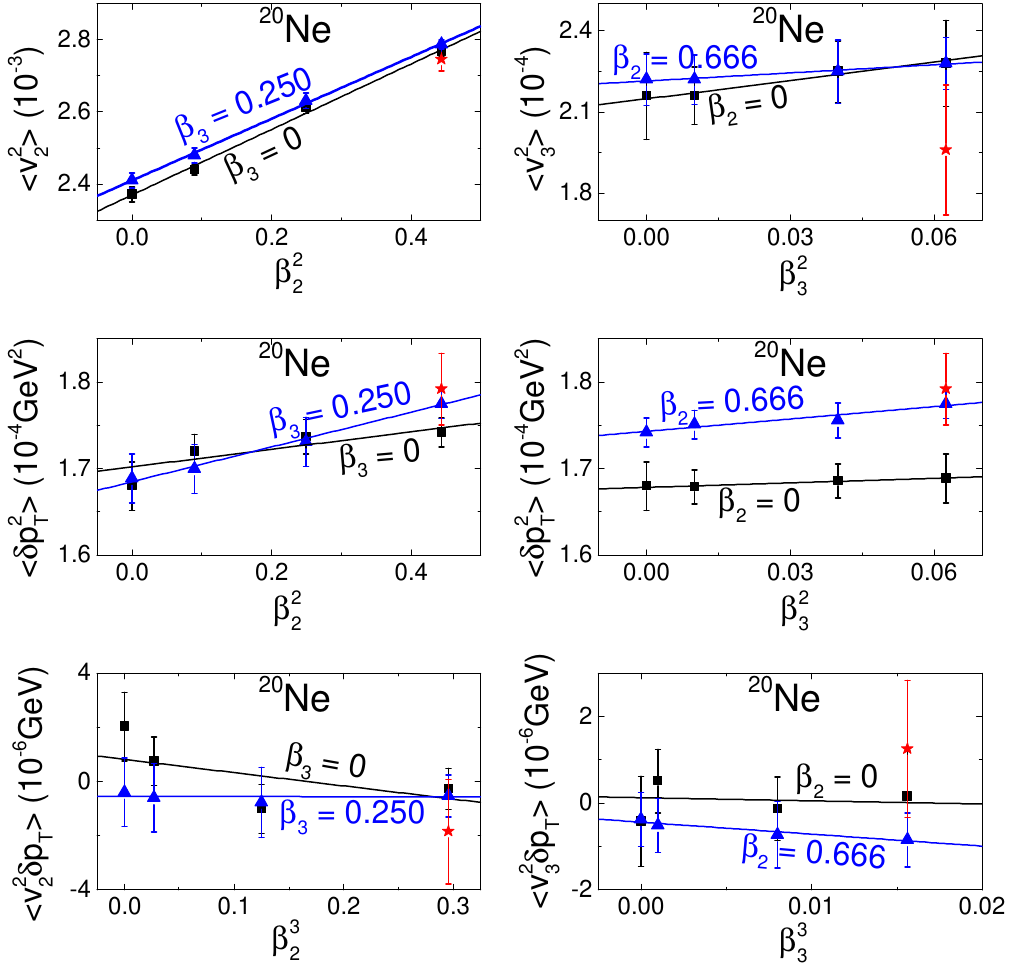}\includegraphics[scale=0.45]{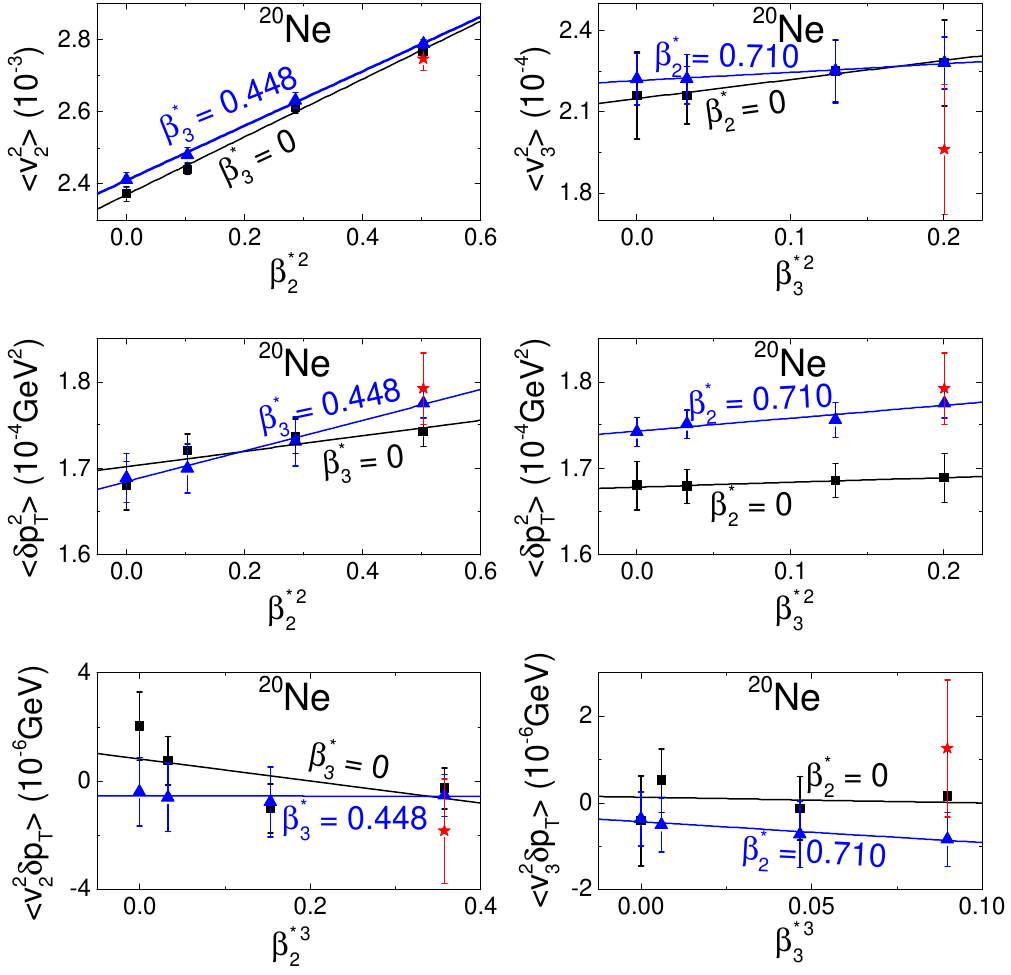}
  \caption{Relations between $\langle v_{2(3)}^2 \rangle$ and $\beta_{2(3)}^2$ (first row), $\langle \delta p_T^2 \rangle$ and $\beta_{2(3)}^2$ (second row), as well as $\langle v_{2(3)}^2\delta p_T \rangle$ and $\beta_{2(3)}^3$ (third row) in central ($0-5\%$) $^{20}$Ne+$^{20}$Ne collisions at $\sqrt{s_{NN}}=200$ GeV. Black squares and blue triangles represent results from density distributions of the WS form with different deformation parameters, while red stars represent results from realistic density distributions with $\alpha$ clusters. The third and fourth columns show similar relations but with $\beta^\star_{2(3)}$.}
  \label{Ne20}
\end{figure*}

Now we investigate whether the deformation probes shown in the previous subsection work well for collisions of light nuclei with $\alpha$ clusters, and we perform AMPT simulations for central collisions of $^{12}$C+$^{12}$C, $^{16}$O+$^{16}$O, $^{20}$Ne+$^{20}$Ne, and $^{10}$Be+$^{10}$Be also at $\sqrt{s_{NN}}=200$ GeV. In order to carry out a fair comparison, we construct an axial symmetric deformed WS density distribution as Eq.~(\ref{dfms}) for each configuration of light nuclei, with $\beta_2$ and $\beta_3$ in the deformed WS distribution adjusted to reproduce $\beta_2^\star$ and $\beta_3^\star$ calculated using Eq.~(\ref{betastar}) from the realistic density distribution with $\alpha$ clusters. The values of the radius parameter $R_0$ and the diffuseness parameter $a$ in the deformed WS distribution are determined in such a way that the values of $\langle r^2 \rangle$ and $\langle r^4 \rangle$ should be the same as those from the realistic density distribution with $\alpha$ clusters, where the $l$th-order moment of $r$ is defined as $\langle r^l \rangle = \int \rho({\vec{r}}) r^l d^3 {r}/\int \rho({\vec{r}}) d^3 {r}$.

Figure~\ref{rho} compares the realistic density distributions for $^{12}$C, $^{16}$O, $^{20}$Ne, and $^{10}$Be with different $\alpha$-cluster configurations as shown in Fig.~\ref{conf} and the deformed WS distributions with parameters determined by the realistic density distributions as described above. The values of $\beta_2^\star$ and $\beta_3^\star$ from realistic density distributions as well as those of $\beta_2$ and $\beta_3$ used in the parameterized deformed WS distribution for different cases are given in Table~\ref{T1}. Here the density distributions of the deformed WS form are axial symmetric with respective to the $z$ axis. For the density distributions with realistic $\alpha$-cluster structures, they are also axial symmetric with respective to the $z$ axis, except the triangle-shaped $^{12}$C and tetrahedron-shaped $^{16}$O, for which the density distributions are plotted in the $3-\alpha$ plane for a better vision. Obviously, the deformed WS distributions are quite different from the realistic ones with $\alpha$ clusters in most cases, and one may expect that they may lead to different values of the probes even if they have the same $\beta_2^\star$ and $\beta_3^\star$.

Figure~\ref{lightnuclei} displays how the deformation probes using deformed WS density distributions deviate from those using realistic density distributions with different $\alpha$-cluster configurations for $^{12}$C, $^{16}$O, and $^{10}$Be. The left three columns show relation between probes and $\beta_{2(3)}$ in the deformed WS distribution, and the right three columns show relations between probes and $\beta^\star_{2(3)}$ calculated from Eq.~(\ref{betastar}). For both parameterized WS distributions or realistic density distributions as shown in Fig.~\ref{rho}, these nuclei have either finite $\beta_2$ ($\beta^\star_2$) or finite $\beta_3$ ($\beta^\star_3$). For the WS density distributions, we calculate results from spherical and deformed density distributions with the same $R_0$ and $a$, as shown by black squares in Fig.~\ref{lightnuclei}. Basically, $\langle v_{2(3)}^2 \rangle$, $\langle \delta p_T^2 \rangle$, and $\langle v_{2(3)}^2\delta p_T \rangle$ follow qualitatively linear relations with $\beta_{2(3)}^2$ or $\beta_{2(3)}^3$, consistent with results from $^{96}$Zr as shown in Sec.~\ref{Zr96}, indicated by solid lines in Fig.~\ref{lightnuclei}. However, results from most realistic density distributions with $\alpha$ clusters deviate from these linear relations. The chain structure of $^{12}$C and $^{16}$O as well as the realistic density distribution for $^{10}$Be lead to larger $\langle v_{2}^2 \rangle$, larger $\langle \delta p_T^2 \rangle$, and smaller $\langle v_{2}^2\delta p_T \rangle$, and this can be understood from the larger asymmetries in $z$ and $r_{xy}$ directions for the realistic density distributions compared to those for a deformed WS form as shown in Fig.~\ref{rho}. The triangle structure of $^{12}$C leads to a smaller $\langle v_{3}^2 \rangle$, a larger $\langle \delta p_T^2 \rangle$, and a smaller $\langle v_{3}^2\delta p_T \rangle$, compared to a deformed WS distribution with the same finite $\beta_3$. The smaller $\langle \epsilon_3^2 \rangle$ from the triangle structure of $^{12}$C can be intuitively understood, since the three $\alpha$ clusters actually form a plane, which generally does not lead to a large $\epsilon_3$ of the overlap region for arbitrary collision configurations, while there are more collision configurations for a deformed WS distribution with a finite $\beta_3$ to have a large $\epsilon_3$. The much larger $\langle \delta p_T^2 \rangle$ originating from the larger fluctuation of the inverse overlap's area $\langle \delta d_\perp^2 \rangle$ for the triangle structure of $^{12}$C compared to the corresponding deformed WS distribution can also be understood in the similar way. The difference is smaller for the tetrahedron configuration of $^{16}$O, since the density distributions from a deformed WS form and a realistic calculation are not quite different, as can be seen from Fig.~\ref{rho}. The relations between deformation probes and $\beta_{2(3)}^\star$ show qualitatively similar behaviors except with smaller slopes. Here we don't see a robust deformation probe among those investigated in the present study that is only sensitive to the deformation parameters of colliding nuclei but insensitive to the existence of $\alpha$ clusters.

Figure~\ref{Ne20} shows similar results as Fig.~\ref{lightnuclei} but for $^{20}$Ne which has both finite $\beta_2$ and $\beta_3$. The left two columns show relation between probes and $\beta_{2(3)}$ in the deformed WS distribution, and the right two columns show relations between probes and $\beta^\star_{2(3)}$ calculated from Eq.~(\ref{betastar}). Using the density distribution of the WS form, qualitatively linear relations between $\langle v_{2(3)}^2 \rangle$, $\langle \delta p_T^2 \rangle$, and $\beta_{2(3)}^2$ for a fixed $\beta_{3(2)}$ are observed, as shown by black squares and blue triangles as well as solid lines, and the behaviors are qualitatively consistent with those from $^{96}$Zr+$^{96}$Zr collisions as shown in Sec.~\ref{Zr96}. Again, the relations between deformation probes and $\beta_{2(3)}^\star$ show qualitatively similar behaviors, except with smaller slopes. Overall, for the $5-\alpha$ cluster structure of $^{20}$Ne, we found that the values of the resulting deformation probes are not too different from those obtained from the parameterized WS distribution within statistical error.

\section{conclusions}
\label{summary}

We have investigated how the probes of deformation parameters $\beta_n$ of colliding nuclei in their collisions at relativistic energies, such as anisotropic flows $\langle v_n^2 \rangle$, transverse momentum fluctuations $\langle \delta p_T^2 \rangle$, and their correlations $\langle v_n^2 \delta p_T \rangle$, work for light nuclei with large $\beta_n$ and different $\alpha$-cluster configurations. By assuming a uniform density distribution with a sharp surface, we have derived the relations between the above probes and $\beta_n$, where higher-order relations and cross relations are observed. The performance of these probes is also investigated with AMPT simulations of collisions of heavy nuclei by assuming they have large $\beta_n$. While the linear relations between $\beta^2_n$, $\langle v_n^2 \rangle$, and $\langle \delta p_T^2 \rangle$ and that between $\beta^3_n$ and $\langle v_n^2 \delta p_T \rangle$ can be violated for extremely large $\beta_{n}$, they are mostly valid for realistic values of $\beta_n$, as long as the density distribution of colliding nuclei can be described by a deformed WS form. However, using more realistic density distributions with $\alpha$ clusters for light nuclei, these probes can deviate from those using a deformed WS form with the same deformation parameters, and the amount of deviation can be different for different $\alpha$-cluster configurations. For the tetrahadron structure of $^{16}$O and the $5-\alpha$ cluster structure of $^{20}$Ne, it is difficult to distinguish the difference in the deformation probes from realistic density distributions and WS density distributions. Therefore, specific probes for $\alpha$-cluster structures in these nuclei are very much demanded in future analysis. For other cases, no robust deformation probe among those investigated in the present study, which is only sensitive to the $\beta_n$ of colliding nuclei but insensitive to the existence of $\alpha$ clusters, is observed, so care must be taken when one tries to extract the deformation of light nuclei.

\begin{acknowledgments}
This work is supported by the Strategic Priority Research Program of the Chinese Academy of Sciences under Grant No. XDB34030000, the National Natural Science Foundation of China under Grant Nos. 12375125, 12035011, and 11975167, and the Fundamental Research Funds for the Central Universities.
\end{acknowledgments}

\appendix

\section{Relations between initial geometry and deformation}
\label{app}

Here we try to derive the relations between the initial geometry in relativistic heavy-ion collisions at zero impact parameter and the deformation parameters $\beta_n$ of colliding nuclei. The initial geometry is characterized by the $n$th-order anisotropy coefficient $\epsilon_n$, the fluctuation of the overlap's inverse area $\delta d_\perp^2$, and their correlation $\epsilon_n^2 \delta d_\perp$. We start from a density distribution with a general deformed WS form
\begin{equation}
\rho(r,\theta, \phi) = \frac{\rho_0}{1+\exp[(r-R(\theta,\phi))/a]}
\end{equation}
with $R(\theta,\phi)=R_0[1+\sum_{l,m} \beta_l \alpha_{l,m} Y_{l,m}(\theta,\phi)]$, but we take the limit of $a \rightarrow 0$ so that the density distribution is uniform with a sharp surface. We basically follow the procedure in Refs.~\cite{Jia:2021tzt,Jia:2021qyu} but keep more higher-order $\beta_l$ terms.

$\epsilon_n$ with respect to the event plane $\Phi_n$ in central tip-tip relativistic heavy-ion collisions can be formally expressed as
\begin{equation}
\epsilon_n e^{in\Phi_n}=-\frac {\int r^n \sin^n(\theta) e^{in\phi} \rho(r,\theta,\phi)d^3r}{\int r^n \sin^n(\theta)  \rho(r,\theta,\phi)d^3r}.
\end{equation}
By using $Y_n^n = \sqrt{\frac{(2n+1)!!}{4\pi(2n)!!}} \sin^n(\theta) e^{in\phi}$ and uniform density distribution within $r \in [0, R_0(1+\sum_{l,m} \beta_l \alpha_{l,m} Y_{l,m})]$, the above equation can be further written as
\begin{widetext}
\begin{eqnarray}
\epsilon_n e^{in\Phi_n} &=& -\sqrt{\frac{4\pi (2n)!!}{(2n+1)!!}}\frac {\int (1+\sum_{l,m} \beta_l\alpha_{l,m}Y_l^m)^{n+3}Y_n^n \sin(\theta) d\theta d\phi}{\int (1+\sum_{l,m} \beta_l\alpha_{l,m}Y_l^m)^{n+3} \sin^{n+1}(\theta) d\theta d\phi} \notag\\
&\approx& -\sqrt{\frac{4\pi (2n)!!}{(2n+1)!!}} \frac{\int [1+(n+3) \sum_{l,m} \beta_l\alpha_{l,m}Y_l^m+\frac{(n+3)(n+2)}{2} \left(\sum_{l,m} \beta_l\alpha_{l,m}Y_l^m\right)^2]Y_n^n \sin(\theta) d\theta d\phi}{\int [1+(n+3)\sum_{l,m} \beta_l\alpha_{l,m}Y_l^m] \sin^{n+1}(\theta) d\theta d\phi}.
\end{eqnarray}
It is seen that we consider additional $\beta_l^2$ terms in the expansion of the numerator, compared to Ref.~\cite{Jia:2021tzt}. By defining
\begin{equation}
A_n=\frac{(n+3)\Gamma(1+1/2+n/2)}{\pi \Gamma(1+n/2)}\sqrt{\frac{(2n)!!}{(2n+1)!!}}
\end{equation}
and
\begin{equation}
B_n=\frac{(n+3)(n+2)\Gamma(1+1/2+n/2)}{2\pi \Gamma(1+n/2)}\sqrt{\frac{(2n)!!}{(2n+1)!!}}
\end{equation}
and by rotating the two nuclei with the same Euler angles $\Omega=(\alpha,\beta,\gamma)$, the above equation can be further written as
\begin{eqnarray}
\epsilon_n e^{in\Phi_n} = &&-A_n\beta _n\sum_m \alpha_{n,m}D_{n,m}^n-B_n \int \sum_{l_1,m_1}\beta_{l_1}\alpha_{l_1,m_1}Y_{l_1}^{m_1}\sum_{l_2,m_2}\beta_{l_2}\alpha_{l_2,m_2}Y_{l_2}^{m_2}\sum_{m'}D_{n,m'}^n Y_n^{m'}  \sin(\theta) d\theta d\phi \notag\\
= &&-A_n\beta _n\sum_m \alpha_{n,m}D_{n,m}^n \notag\\
&&- B_n \sum_{l_1,m_1,l_2,m_2,m'}\left(\beta_{l_1}\beta_{l_2}\alpha_{l_1,m_1}\alpha_{l_2,m_2} D_{n,m'}^n\sqrt{\frac{(2l_1+1)(2l_2+1)(2n+1)}{4\pi}}\left(\begin {array}{ccc}l_1&l_2&n\\0&0&0\end {array}\right) \left(\begin {array}{ccc}l_1&l_2&n\\ m_1&m_2&m'\end {array} \right)\right), \notag\\
\end{eqnarray}
with $D_{n,m}^n(\alpha,\beta,\gamma)$ being the Wigner rotation matrix, and $\left(\begin {array}{ccc}j_1&j_2&j\\ m_1&m_2&m\end {array} \right)$ representing the Wigner 3j-Symbol. Taking the square of the above expression leads to
\begin{eqnarray}
\epsilon_n^2&=&A_n^2\beta _n^2\left(\sum_m \alpha_{n,m}D_{n,m}^n\right)\left(\sum_{m'} \alpha_{n,m'}D_{n,m'}^n\right)\notag\\
&+&2A_nB_n\beta _n \sum_{l_1,m_1,l_2,m_2,m',m}\beta_{l_1}\beta_{l_2} \alpha_{n,m}D_{n,m}^n\alpha_{l_1,m_1}\alpha_{l_2,m_2} D_{n,m'}^n\sqrt{\frac{(2l_1+1)(2l_2+1)(2n+1)}{4\pi}}\left(\begin {array}{ccc}l_1&l_2&n\\0&0&0\end {array}\right) \left(\begin {array}{ccc}l_1&l_2&n\\ m_1&m_2&m'\end {array}\right) \notag\\
&+&O(\beta_n^4).
\end{eqnarray}
In the present study, we only use finite $\beta_2$ and $\beta_3$ but $\beta_n=0$ for $n>3$. The event average here is equal to the rotational average $\int (...) d\Omega /(8\pi^2)=\int (...) d\alpha \sin(\beta) d\beta d\gamma/(8\pi^2)$, and then we can get numerically the relations as Eqs.~(\ref{ep2}) and (\ref{ep3}) in the axial symmetric case. For independent rotation of the two nuclei, as shown in Ref.~\cite{Jia:2021tzt}, qualitatively similar relations apply except that the coefficients are a factor of 2 smaller.

Next, we calculate the fluctuation $\delta d_\perp^2$ of the overlap's inverse area $d_\perp = 1/\sqrt{\overline{x^2}~\overline{y^2}}$. At zero impact parameter, $\overline{x^2}$ in tip-tip collisions can be formally expressed as
\begin{eqnarray}
\overline{x^2} &=& \frac {\int r^2 \sin^2(\theta) \cos^2(\phi) \rho(r,\theta, \phi)d^3r}{\int\rho(r,\theta, \phi)d^3r} \notag\\
&=&\frac {3R_0^2}{20\pi} \int (1+\sum_{l,m} \beta_l\alpha_{l,m}Y_l^m)^5 \left[\frac 1 3 -\frac 2 3 \sqrt{\frac {\pi} 5}Y_2^0+\sqrt{\frac {2\pi}{15}}(Y_2^2+Y_2^{-2})\right] \sin(\theta) d\theta d\phi \notag\\
&\approx& \frac {3R_0^2}{20\pi}\int \left(1+5\sum_{l,m} \beta_l\alpha_{l,m}Y_l^m+10\sum_{l_1,m_1,l_2,m_2} \beta_{l_1}\beta_{l_2}\alpha_{l_1,m_1}\alpha_{l_2,m_2}Y_{l_1}^{m_1} Y_{l_2}^{m_2}\right) \left[\frac 1 3 -\frac 2 3 \sqrt{\frac {\pi} 5}Y_2^0+\sqrt{\frac {2\pi}{15}}(Y_2^2+Y_2^{-2})\right] \sin(\theta) d\theta d\phi \notag\\
&=& \frac{R_0^2}5+\frac {3R_0^2}{20\pi}\int\left[{ -\frac{10}3\sqrt{\frac {\pi} 5} \sum_{l,m} \beta_l\alpha_{l,m}Y_l^mY_2^0+5\sqrt{\frac {2\pi}{15}}\sum_{l,m} \beta_l\alpha_{l,m}Y_l^m(Y_2^2+Y_2^{-2})+\frac{10}3\sum_{l_1,m_1,l_2,m_2} \beta_{l_1}\beta_{l_2}\alpha_{l_1,m_1}\alpha_{l_2,m_2}Y_{l_1}^{m_1} Y_{l_2}^{m_2} }\right. \notag\\
&&\left.{ -\frac {20} 3 \sqrt{\frac {\pi} 5}\sum_{l_1,m_1,l_2,m_2} \beta_{l_1}\beta_{l_2}\alpha_{l_1,m_1}\alpha_{l_2,m_2}Y_{l_1}^{m_1} Y_{l_2}^{m_2}Y_2^0  +10\sqrt{\frac {2\pi}{15}}\sum_{l_1,m_1,l_2,m_2} \beta_{l_1}\beta_{l_2}\alpha_{l_1,m_1}\alpha_{l_2,m_2}Y_{l_1}^{m_1} Y_{l_2}^{m_2}(Y_2^2+Y_2^{-2}) }\right]\sin(\theta) d\theta d\phi. \notag\\
\end{eqnarray}
In the above, the relation $\cos^2(\phi) \sin^2(\theta)=\frac 1 3 -\frac 2 3 \sqrt{\frac {\pi} 5}Y_2^0+\sqrt{\frac {2\pi}{15}}(Y_2^2+Y_2^{-2})$ is used, and we keep higher-order $\beta_l$ terms compared to Ref.~\cite{Jia:2021qyu}. By rotating the two nuclei with the same Euler angles $\Omega$ and carrying out the integral of the spherical harmonics, the above equation can be further written as
\begin{eqnarray}
\overline{x^2} &=& \frac{R_0^2}5+\frac {3R_0^2}{20\pi}\left[{ -\frac{10}3\sqrt{\frac {\pi} 5} \sum_{m} \beta_2\alpha_{2,m}D_{0,m}^2+5\sqrt{\frac {2\pi}{15}}\sum_{m} \beta_2\alpha_{2,m}(D_{2,m}^2+D_{-2,m}^2)+\frac{10}3\sum_{l,m,} \beta_l^2\alpha_{l,m}^2 }\right.\notag\\
&&\left.{ -\frac {20} 3 \sqrt{\frac {\pi} 5}\sum_{l_1,m_1,l_2,m_2,m} \beta_{l_1}\beta_{l_2}\alpha_{l_1,m_1}\alpha_{l_2,m_2}D_{0,m}^2\sqrt{\frac{5(2l_1+1)(2l_2+1)}{4\pi}}\left(\begin {array}{ccc}l_1&l_2&2\\0&0&0\end {array}\right)\left(\begin {array}{ccc}l_1&l_2&2\\ m_1&m_2&m\end {array} \right) } \right.\notag\\
&&\left.{+10\sqrt{\frac {2\pi}{15}}\sum_{l_1,m_1,l_2,m_2,m} \beta_{l_1}\beta_{l_2}\alpha_{l_1,m_1}\alpha_{l_2,m_2}(D_{2,m}^2+D_{-2,m}^2)\sqrt{\frac{5(2l_1+1)(2l_2+1)}{4\pi}}\left(\begin {array}{ccc}l_1&l_2&2\\0&0&0\end {array}\right)\left(\begin {array}{ccc}l_1&l_2&2\\ m_1&m_2&m\end {array} \right) }\right]\notag\\
&=& \frac{R_0^2}5+\frac {R_0^2}{\sqrt{20\pi}}\sum_{m} \beta_2\alpha_{2,m}\left[-D_{0,m}^2+\sqrt{\frac 3 2 }(D_{2,m}^2+D_{-2,m}^2)\right]+\frac{R_0^2}{2\pi}\sum_{l,m} \beta_l^2\alpha_{l,m}^2 \notag\\
&&+\frac {R_0^2}{\sqrt{5\pi}}\sum_{l_1,m_1,l_2,m_2,m} \beta_{l_1}\beta_{l_2}\alpha_{l_1,m_1}\alpha_{l_2,m_2}\sqrt{\frac{5(2l_1+1)(2l_2+1)}{4\pi}}\left(\begin {array}{ccc}l_1&l_2&2\\0&0&0\end {array}\right)\left(\begin {array}{ccc}l_1&l_2&2\\ m_1&m_2&m\end {array} \right)\left[-D_{0,m}^2+\sqrt{\frac 3 2 }(D_{2,m}^2+D_{-2,m}^2)\right].\notag\\
\end{eqnarray}
Similarly, we can formally express $\overline{y^2}$ in central tip-tip collisions as
\begin{eqnarray}
\overline{y^2}&=&\frac {\int r^2 \sin^2(\theta) \sin^2(\phi) \rho(r,\theta,\phi)d^3r}{\int\rho(r,\theta,\phi)d^3r} \notag\\
&\approx& \frac{R_0^2}5+\frac {R_0^2}{\sqrt{20\pi}}\sum_{m} \beta_2\alpha_{2,m}\left[-D_{0,m}^2-\sqrt{\frac 3 2 }(D_{2,m}^2+D_{-2,m}^2)\right]+\frac{R_0^2}{2\pi}\sum_{l,m} \beta_l^2\alpha_{l,m}^2 \notag\\
&&+\frac {R_0^2}{\sqrt{5\pi}}\sum_{l_1,m_1,l_2,m_2,m} \beta_{l_1}\beta_{l_2}\alpha_{l_1,m_1}\alpha_{l_2,m_2}\sqrt{\frac{5(2l_1+1)(2l_2+1)}{4\pi}}\left(\begin {array}{ccc}l_1&l_2&2\\0&0&0\end {array}\right)\left(\begin {array}{ccc}l_1&l_2&2\\ m_1&m_2&m\end {array} \right)\left[-D_{0,m}^2-\sqrt{\frac 3 2 }(D_{2,m}^2+D_{-2,m}^2)\right], \notag\\
\end{eqnarray}
for which we have used the relation $\sin^2(\phi) \sin^2(\theta)=\frac 1 3 -\frac 2 3 \sqrt{\frac {\pi} 5}Y_2^0-\sqrt{\frac {2\pi}{15}}(Y_2^2+Y_2^{-2})$ and keep higher-order $\beta_l$ terms compared to Ref.~\cite{Jia:2021qyu}. Thus, up to the $\beta_l^3$ term, $\overline{x^2}~\overline{y^2}$ can be written as
\begin{eqnarray}
\overline{x^2}~\overline{y^2}&=&\frac{R_0^4}{25}-\frac {R_0^4}{5\sqrt{5\pi}}\sum_{m} \beta_2\alpha_{2,m}D_{0,m}^2+\frac {R_0^4}{20\pi}\sum_{m_1,m_2}\beta_2^2\alpha_{2,m_1}\alpha_{2,m_2}\left[D_{0,m_1}^2D_{0,m_2}^2-\frac 3 2 (D_{2,m_1}^2+D_{-2,m_1}^2)(D_{2,m_2}^2+D_{-2,m_2}^2)\right] \notag\\
&&+\frac{R_0^4}{5\pi}\sum_{l,m} \beta_l^2\alpha_{l,m}^2-\frac{R_0^4}{\sqrt{20\pi^3}}\sum_{l,m_1,m_2}\beta_2\beta_l^2\alpha_{2,m_1}\alpha_{l,m_2}^2D_{0,m_1}^2  \notag\\
&&-\frac {2R_0^4}{5\sqrt{5\pi}}\sum_{l_1,m_1,l_2,m_2,m} \beta_{l_1}\beta_{l_2}\alpha_{l_1,m_1}\alpha_{l_2,m_2}\sqrt{\frac{5(2l_1+1)(2l_2+1)}{4\pi}}\left(\begin {array}{ccc}l_1&l_2&2\\0&0&0\end {array}\right)\left(\begin {array}{ccc}l_1&l_2&2\\ m_1&m_2&m\end {array} \right)D_{0,m}^2\notag\\
&&+\frac{R_0^4}{5\pi}\sum_{l_1,m_1,l_2,m_2,m,m'} \beta_2\beta_{l_1}\beta_{l_2}\alpha_{l_1,m_1}\alpha_{l_2,m_2}\alpha_{2,m'}\sqrt{\frac{5(2l_1+1)(2l_2+1)}{4\pi}}\left(\begin {array}{ccc}l_1&l_2&2\\0&0&0\end {array}\right)\left(\begin {array}{ccc}l_1&l_2&2\\ m_1&m_2&m\end {array} \right)\notag\\
&&\times\left[D_{0,m}^2D_{0,m'}^2-\frac 3 2 (D_{2,m}^2+D_{-2,m}^2)(D_{2,m'}^2+D_{-2,m'}^2)\right]+O(\beta_l^4).
\end{eqnarray}
By definition $d_\perp = 1/\sqrt{\overline{x^2}~\overline{y^2}}$ can then be expressed as
\begin{eqnarray}
d_\perp&\approx&\frac5{R_0^2} \left\{ {1+\sqrt{\frac5 {4\pi}}\sum_{m} \beta_2\alpha_{2,m}D_{0,m}^2-\frac 5{8\pi}\sum_{m_1,m_2}\beta_2^2\alpha_{2,m_1}\alpha_{2,m_2}\left[D_{0,m_1}^2D_{0,m_2}^2-\frac 3 2 (D_{2,m_1}^2+D_{-2,m_1}^2)(D_{2,m_2}^2+D_{-2,m_2}^2)\right]} \right. \notag\\
&&\left.-\frac5{2\pi}\sum_{l,m} \beta_l^2\alpha_{l,m}^2+\sqrt{\frac5 \pi}\sum_{l_1,m_1,l_2,m_2,m} \beta_{l_1}\beta_{l_2}\alpha_{l_1,m_1}\alpha_{l_2,m_2}\sqrt{\frac{5(2l_1+1)(2l_2+1)}{4\pi}}\left(\begin {array}{ccc}l_1&l_2&2\\0&0&0\end {array}\right)\left(\begin {array}{ccc}l_1&l_2&2\\ m_1&m_2&m\end {array} \right)D_{0,m}^2 \right.\notag\\
&&\left.+\sqrt{\frac{125}{16\pi^3}}\sum_{l,m_1,m_2}\beta_2\beta_l^2\alpha_{2,m_1}\alpha_{l,m_2}^2D_{0,m_1}^2-\frac{5}{2\pi}\sum_{l_1,m_1,l_2,m_2,m,m'} \beta_2\beta_{l_1}\beta_{l_2}\alpha_{l_1,m_1}\alpha_{l_2,m_2}\alpha_{2,m'}\right. \notag\\
&&\left.\times\sqrt{\frac{5(2l_1+1)(2l_2+1)}{4\pi}}\left(\begin {array}{ccc}l_1&l_2&2\\0&0&0\end {array}\right)\left(\begin {array}{ccc}l_1&l_2&2\\ m_1&m_2&m\end {array} \right)\left[D_{0,m}^2D_{0,m'}^2 -\frac 3 2 (D_{2,m}^2+D_{-2,m}^2)(D_{2,m'}^2+D_{-2,m'}^2)\right] \right\}+O(\beta_l^4).\notag\\
\end{eqnarray}
Let's define the deviation of a quantity $A$ away from its rotational average value as $\delta A = A - \langle A \rangle$, where $\langle A \rangle=\int A d\Omega/(8\pi)^2$. In this way, $\delta d_{\perp}$, $(\delta d_{\perp})^2$, and $\epsilon_n^2 \delta d_{\perp}$ can then be formally expressed as
\begin{eqnarray}
\delta d_{\perp}&=&\frac5{R_0^2}\left\{ {\sqrt{\frac5 {4\pi}}\sum_{m} \beta_2\alpha_{2,m}\delta D_{0,m}^2-\frac 5{8\pi}\sum_{m_1,m_2}\beta_2^2\alpha_{2,m_1}\alpha_{2,m_2}\delta\left[D_{0,m_1}^2D_{0,m_2}^2-\frac 3 2 (D_{2,m_1}^2+D_{-2,m_1}^2)(D_{2,m_2}^2+D_{-2,m_2}^2)\right]}\right. \notag\\
&&\left.{+\sqrt{\frac5 \pi}\sum_{l_1,m_1,l_2,m_2,m} \beta_{l_1}\beta_{l_2}\alpha_{l_1,m_1}\alpha_{l_2,m_2}\sqrt{\frac{5(2l_1+1)(2l_2+1)}{4\pi}}\left(\begin {array}{ccc}l_1&l_2&2\\0&0&0\end {array}\right)\left(\begin {array}{ccc}l_1&l_2&2\\ m_1&m_2&m\end {array} \right)\delta D_{0,m}^2}\right.\notag\\
&&\left.{+\sqrt{\frac{125}{16\pi^3}}\sum_{l,m_1,m_2}\beta_2\beta_l^2\alpha_{2,m_1}\alpha_{l,m_2}^2\delta D_{0,m_1}^2-\frac{5}{2\pi}\sum_{l_1,m_1,l_2,m_2,m,m'} \beta_2\beta_{l_1}\beta_{l_2}\alpha_{l_1,m_1}\alpha_{l_2,m_2}\alpha_{2,m'} }\right. \notag\\
&&\left.{\times\sqrt{\frac{5(2l_1+1)(2l_2+1)}{4\pi}}\left(\begin {array}{ccc}l_1&l_2&2\\0&0&0\end {array}\right)\left(\begin {array}{ccc}l_1&l_2&2\\ m_1&m_2&m\end {array} \right)\delta\left[D_{0,m}^2D_{0,m'}^2-\frac 3 2 (D_{2,m}^2+D_{-2,m}^2)(D_{2,m'}^2+D_{-2,m'}^2)\right]} \right\}+O(\beta_l^4),\notag\\
\end{eqnarray}
\begin{eqnarray}
(\delta d_{\perp})^2&=&\frac{25}{R_0^4}\left\{{\frac5 {4\pi}\sum_{m_1,m_2} \beta_2^2\alpha_{2,m_1}\alpha_{2,m_2}\delta D_{0,m_1}^2\delta D_{0,m_2}^2-\sqrt{\frac {125}{64\pi^3}}\sum_{m,m_1,m_2}\beta_2^3\alpha_{2,m}\alpha_{2,m_1}\alpha_{2,m_2}}\right.\notag\\
&&\left.{\times\delta D_{0,m}^2\delta\left[D_{0,m_1}^2D_{0,m_2}^2-\frac 3 2 (D_{2,m_1}^2+D_{-2,m_1}^2)(D_{2,m_2}^2+D_{-2,m_2}^2)\right]}\right.\notag\\
&&\left.{+\frac5 \pi \beta_2\sum_{l_1,m_1,l_2,m_2,m,m'} \beta_{l_1}\beta_{l_2}\alpha_{l_1,m_1}\alpha_{l_2,m_2}\alpha_{2,m'}\sqrt{\frac{5(2l_1+1)(2l_2+1)}{4\pi}}\left(\begin {array}{ccc}l_1&l_2&2\\0&0&0\end {array}\right)\left(\begin {array}{ccc}l_1&l_2&2\\ m_1&m_2&m\end {array} \right)\delta D_{0,m}^2\delta D_{0,m'}^2} \right\}+O(\beta_l^4),\notag\\
\end{eqnarray}
and
\begin{eqnarray}
\epsilon_n^2 \delta d_\perp&=&\frac 5{R_0^2}A_n^2\left \{ {\sqrt{\frac5 {4\pi}} \beta_2\beta_n^2\sum_{m_1,m_2,m_3}\alpha_{2,m_1}\alpha_{n,m_2}\alpha_{n,m_3}D_{n,m_2}^n D_{n,m_3}^n\delta D_{0,m_1}^2}\right.\notag\\
&&\left.{-\frac 5{8\pi}\beta_2^2\beta_n^2\sum_{m_1,m_2,m_3,m_4}\alpha_{2,m_1}\alpha_{2,m_2}\alpha_{n,m_3}\alpha_{n,m_4}\delta\left[D_{0,m_1}^2D_{0,m_2}^2-\frac 3 2 (D_{2,m_1}^2+D_{-2,m_1}^2)(D_{2,m_2}^2+D_{-2,m_2}^2)\right]D_{n,m_3}^n D_{n,m_4}^n}\right.\notag\\
&&\left.{+\sqrt{\frac5 \pi}\beta_n^2\sum_{l_1,m_1,l_2,m_2,m_3,m_4,m} \beta_{l_1}\beta_{l_2}\alpha_{l_1,m_1}\alpha_{l_2,m_2}\alpha_{n,m_3}\alpha_{n,m_4}}\right. \notag\\
&&\left.{\times\sqrt{\frac{5(2l_1+1)(2l_2+1)}{4\pi}}\left(\begin {array}{ccc}l_1&l_2&2\\0&0&0\end {array}\right)\left(\begin {array}{ccc}l_1&l_2&2\\ m_1&m_2&m\end {array} \right)D_{n,m_3}^n D_{n,m_4}^n\delta D_{0,m}^2}\right.\notag\\
&&\left.{+\sqrt{\frac5 {4\pi}}(n+2) \beta_2\beta_n\sum_{l_1,m_1,l_2,m_2,m_3,m',m}\beta_{l_1}\beta_{l_2} \alpha_{n,m}\alpha_{l_1,m_1}\alpha_{l_2,m_2} \alpha_{2,m_3} }\right. \notag\\
&&\left.{\times\sqrt{\frac{(2l_1+1)(2l_2+1)(2n+1)}{4\pi}}\left(\begin {array}{ccc}l_1&l_2&n\\0&0&0\end {array}\right) \left(\begin {array}{ccc}l_1&l_2&n\\ m_1&m_2&m'\end {array}\right)D_{n,m}^nD_{n,m'}^n\delta D_{0,m_3}^2}\right \}+O(\beta_l^5),
\end{eqnarray}
respectively.
Again, using finite $\beta_2$ and $\beta_3$ but $\beta_n=0$ for $n>3$ and taking the rotational average by integrating over the Euler angles, we can get numerically the relations as Eqs.~(\ref{dd2}), (\ref{ep2dd2}), and (\ref{ep3dd2}) in the axial symmetric case. For independent rotation of the two nuclei, as shown in Ref.~\cite{Jia:2021qyu}, qualitatively similar relations apply except that the coefficients are a factor of 2 smaller.
\end{widetext}
\bibliography{light_nuclei_clusters}
\end{document}